\def\draftversion{false}

\RequirePackage{ifthen}
\ifthenelse{\equal{\draftversion}{false}}{
  \documentclass[citeautoscript,floatfix,aps,prx,twocolumn,preprintnumbers,
      superscriptaddress,longbibliography,amsmath,amssymb]{revtex4-2}
}{
  \documentclass[citeautoscript,floatfix,aps,prx,twocolumn,
      superscriptaddress,longbibliography]{revtex4-2}
}

\usepackage{amsmath}
\usepackage{graphicx}
\usepackage{epstopdf}
\usepackage{natbib}
\usepackage{array}
\usepackage{dcolumn}
\usepackage{bm}
\usepackage{blkarray}
\usepackage{soul}  

\usepackage[usenames,dvipsnames]{color}

\newcommand{\blue}{\textcolor{Black}}
\newcommand\T{\rule{0pt}{2.6ex}}              
\newcommand\B{\rule[-1.2ex]{0pt}{0pt}}        

\soulregister\cite7

\ifthenelse{\equal{\draftversion}{true}}{
  \marginparwidth 2.7in
  \marginparsep 0.5in
  \newcounter{comm} 
  \def\commnext{\stepcounter{comm}}
  \def\commtext{{\bf\color{blue}[\arabic{comm}]}}
  \def\commmar{{\bf\color{blue}[\arabic{comm}]}}
  \def\dvm#1{\commnext\marginpar{\small DV\commmar: #1}\commtext}
  \def\cdm#1{\commnext\marginpar{\small CED\commmar: #1}\commtext}
  \def\msm#1{\commnext\marginpar{\small MS\commmar: #1}\commtext}
  \def\asm#1{\commnext\marginpar{\small AS\commmar: #1}\commtext}
  \def\miq#1{\commnext\marginpar{\small MR\commmar: #1}\commtext}
  \def\mlab#1{\marginpar{\small\bf #1}}
  
}{
  \def\dvm#1{}
  \def\cdm#1{}
  \def\msm#1{}
  \def\asm#1{}
  \def\miq#1{}
  \def\mlab#1{}
  
}

\newcommand{\w}{\omega}
\newcommand{\la}{\lambda}
\newcommand{\e}{\epsilon}

\begin{document}

\title{Dynamical response of noncollinear spin systems at
constrained magnetic moments}

\author{Miquel Royo}
\affiliation{Institut de Ci\`encia de Materials de Barcelona 
(ICMAB-CSIC), Campus UAB, 08193 Bellaterra, Spain}
\email{mroyo@icmab.es}

\author{Massimiliano Stengel}
\affiliation{Institut de Ci\`encia de Materials de Barcelona 
(ICMAB-CSIC), Campus UAB, 08193 Bellaterra, Spain}
\affiliation{ICREA - Instituci\'o Catalana de Recerca i Estudis Avan\c{c}ats, 08010 Barcelona, Spain}
\email{mstengel@icmab.es}

\date{\today}

\begin{abstract}
Noncollinear magnets are notoriously difficult to describe within first-principles approaches based on density-functional theory (DFT) because of the presence of low-lying spin excitations.
At the level of ground-state calculations, several methods exist to constrain the magnetic moments to a predetermined configuration, and thereby accelerate convergence towards self-consistency. Their use in a perturbative context, however, remains very limited.
Here we present a general methodological framework to achieve parametric control over the local spin moments at the linear-response level.
Our strategy builds on the concept of Legendre transform to switch between various flavors of magnetic functionals,
and to relate their second derivatives via simple linear-algebra operations.
Thereby, we can address an arbitrary response function at the time-dependent DFT level of theory with optimal accuracy and minimal computational effort.
In the low frequency limit, we identify the leading correction to the existing adiabatic formulation of the problem [S. Ren \emph{et al.}, Phys. Rev. X {\bf 14}, 011041 (2024)], consisting in a renormalization of the phonon and magnon masses due to electron inertia.
As a demonstration, we apply our methodology to the THz optical response of bulk CrI$_3$ and Cr$_2$O$_3$, where we identify contributions from hybrid (electro)magnons with mixed spin-lattice character.

\end{abstract}

\pacs{71.15.-m, 
        63.20.dk} 
\maketitle

\section{Introduction}

\blue{In recent years, the terahertz (THz) dynamics of magnetically ordered crystals has gained considerable attention due to the potential for creating and manipulating spin-wave excitations through ultrafast laser pulses.~\cite{salikhov2023,hortensius2021,zhang2024}
The research activity in this area has increased at a fast pace after
the breakthrough discovery~\cite{pimenov2006,ValdesPRB07,Sushkov-07,ValdesPRL09} of electromagnons in rare-earth
manganites, 35 years after their theoretical prediction~\cite{baryakhtar1970}.
Since then, a wide range of emergent phenomena have been reported, such as
coherent spin switching~\cite{Schlauderer2019}, magnetic order melting~\cite{Forst2015},
and magnetoelectric coupling~\cite{TakahashiNatPhys12,bilyk2025l},  offering promising opportunities in low-power computing technologies~\cite{barman2021}.
A key challenge in this field consists in identifying the most efficient
microscopic mechanisms that enable optical control of spin-wave excitations,
and in quantitatively understanding  how the ensuing dynamics manifests itself in optical probes (e.g., Faraday rotation).}

\blue{Among the most promising routes that are currently being explored, spin-lattice couplings occupy a central place.
Infrared-active phonons can act as a proxy by converting the electrical component of the external THz radiation into magnetic torques,
and hence endow spin waves with a sizeable dipolar strength~\cite{nova2016,dornes2019,stupakiewicz2021,afanasiev2021}.
Such multimode interaction can also enter a nonlinear regime under intense laser pulses~\cite{Mashkovich21,MetzgerNatCom24,bilyk2025l}, opening
new opportunities for the coherent control of magnetism on sub-picosecond timescales.
However, the underlying mechanisms remain poorly understood at a fundamental level, due to the
conceptual and technical intricacies that the coupled dynamics of the spin-phonon system entails.
Perturbative density-functional calculations are, in principle, ideally suited to tackle these challenges; however, their reliability still depends critically on overcoming fundamental limitations in the state-of-the-art formalism and implementations.}

Since its inception in the 1980s~\cite{zein-84,baroni-87},
density-functional perturbation theory (DFPT) has been widely employed to study the linear response of crystalline systems to perturbations such as phonons~\cite{Baroni-01}, electric fields~\cite{gonze,gonze/lee}, and strains~\cite{hamann-metric}.
More recently, DFPT has been extended to investigate the response of noncollinear magnetic systems to electric fields and phonons, both at the Brillouin zone center~\cite{RicciPRB19} and for arbitrary wave vectors~\cite{UrruPRB19}.
Even earlier, the Sternheimer equation, a cornerstone of DFPT, had been applied in a noncollinear and dynamic (time- or frequency-dependent) regime to calculate spin-wave excitation spectra~\cite{SavrasovPRL98}.
This approach was further refined and optimized in Refs.~\onlinecite{CaoPRB18,GorniEPJB18,LiuPRB23} within various open-source density-functional suites.
Of particular significance is the work by Gorni \emph{et al.}~\cite{GorniEPJB18} who developed a powerful Lanczos chain algorithm to obtain the spin susceptibility over a dense grid of frequencies in a single-shot calculation.
These advancements highlight the potential of DFPT within a frequency-dependent framework as a valuable tool, competitive with many-body methods~\cite{karlsson2000,sasioglu2010,buczek2011,rousseau2012,OlsenPRL21,EsquembrePRB25,LePRB25}.

Recently, promising methods have emerged to treat magnons and phonons on equal footing, emphasizing the importance of such coupled dynamics in accurately evaluating the eigenfrequencies of both excitations~\cite{RenPRX24, BoniniPRL23}.
The approach proposed by Ren \emph{et al.}~\cite{RenPRX24} relies on a generalized adiabatic approximation, where Hessians in the extended parameter space are combined with Berry curvatures, ensuring the correct propagation of time-reversal symmetry breaking from the electronic to the lattice subsystem.
In the clamped-ion limit, this method reduces to the semiclassical approach of Niu and Kleinman~\cite{NiuPRL98} for magnon dynamics, where both Hessians and Berry curvatures can be computed using DFPT, as demonstrated by Lin and Feng~\cite{lin2025}.

A common challenge in all the aforementioned approaches is the difficulty in achieving convergence when calculating the linear response of noncollinear spin systems to external perturbations.
This arises due to magnon resonances, which typically occur at very low energies, significantly increasing the condition number of the linear problem.
These challenges also affect static calculations at the Brillouin zone center, where the acoustic magnon is a Goldstone mode in the absence of spin-orbit coupling, or otherwise has a frequency on the order of a few meV, corresponding to the weak magnetocrystalline anisotropy energy.
More broadly, calculating the spectral function near resonance becomes intractable unless a sufficiently large ``inverse relaxation time'' ($\eta$) is introduced, which severely limits the energy resolution that can be practically achieved.
Furthermore, a unified formulation of DFPT that treats magnons, phonons, and their contributions to the electromagnetic response within a single theoretical and computational framework has yet to be developed.
The current implementation of the method developed by Ren et al.~\cite{RenPRX24} is based on a finite-difference (``frozen-phonon'' and ``frozen-magnon'') calculation of the relevant Hessian and Berry curvatures, and suffers from equally serious numerical convergence issues.
More generally, the adequacy of the adiabatic approximation has only been benchmarked against the "exact" result from nonadiabatic time-dependent DFT at the clamped-ion level~\cite{lin2025}.

In this work, we propose a solution to the convergence issues by constraining the magnetic moments to their unperturbed value during the linear-response calculations.
The constraint, which we introduce via the penalty-function approach, has the main purpose of
stiffening the spin degrees of freedom. This way, we effectively eliminate the problematic magnon resonances from the low-frequency part of the spectrum, and thus achieve an optimally fast convergence of the self-consistent iterations.
Interestingly, our preconditioned functional can be regarded as the magnetic equivalent of the constrained electric displacement (${\bf D}$) method of Ref.~\onlinecite{fixedd}.
Pursuing this analogy, we frame our theory around the concept of Legendre transform, which we use to establish exact relations between the calculated coefficients and the physically relevant (relaxed-spin) response functions.
This way, we can achieve an accurate representation of the coupled spin-phonon resonances while performing the calculation of the relevant coefficients in the nonresonant regime, with a drastic increase in computational efficiency.
In passing, we also establish an exact mapping between our results and those of a hypothetical constrained-DFT calculation based on
the Lagrange multiplier approach.

\blue{We showcase the effectiveness of our approach using bulk
CrI$_3$ and Cr$_2$O$_3$ as testcases of prototypical ferromagnetic and antiferromagnetic insulators, respectively.}
We calculate the frequency-dependent response of \blue{both materials} to atomic displacements,
electric and Zeeman magnetic fields over the frequency range that is relevant to phonons and magnons.
\blue{We focus here on the response at the $\Gamma$ point, while the full Brillouin-zone dispersion will be addressed in a forthcoming publication.}
We then study the nonadiabatic spin-lattice dynamics and discuss its signatures in the magnetic, \blue{(anti)magnetoelectric} and dielectric susceptibilities.
By separately investigating the clamped- and relax-ion regime, we highlight the phonon-(electro)magnon coupling as evidenced by shifts and splittings of resonances, and by  large lattice-mediated spin responses to the external electric field.

Finally, we test the adiabatic formalism of Ren et al.~\cite{RenPRX24} and Lin and Feng~\cite{lin2025}, which emerges as a well-defined approximation to our theory. We find that such an approximation is generally justified, and propose a simple scheme to improve its accuracy in cases where significant deviations exist.
The resulting \emph{second-order adiabatic approximation} differs from the aforementioned schemes by a renormalization of the atomic and magnonic masses. The latter originates from the inertia of the electronic currents that are associated with a given adiabatic path in parameter space.

This work is organized as follows.
In Section II we present
the theoretical framework and main formal results.
In Section III we discuss the
practical implementation within a DFPT framework.
\blue{In Section IV we present the computational parameters and a thorough numerical validation of our approach; this technical Section
may be skipped by nonspecialists on a first read.
In Section V and VI
we present our main physical results for bulk CrI$_3$ and Cr$_2$O$_3$, respectively.}
Finally, in Section VII we present
a brief summary and outlook.

\section{Theory \label{sec:theory}}

In this section, we introduce an efficient first-principles formalism to study the linear response of noncollinear spin systems.
Our approach builds on the penalty-function approach~\cite{MaPRB15} to constrained DFT, which is routinely used in many codes to calculate the ground-state properties of the system in a predetermined magnetic structure. 
Here we focus on the linear-response regime, where we demonstrate that the constraints indeed reduce the condition number of the problem by orders of magnitude, and allow for a robust convergence to the stationary solution. 
Subsequently, the (physically relevant) response of the system in absence of constraints can be recovered exactly at the post-processing stage via a sequence of straightforward linear-algebra operations.

\subsection{Thermodynamic magnetic functionals\label{sec:gsfun}}

We start by defining the magnetization density as~\footnote{Eq.~\eqref{mofr} defines ${\bf m}({\bf r})$ in units of bohr$^{-3}$, and describes the physical magnetization modulo a factor of $-g\mu_{\rm B}/2$. ($g\simeq 2$ is the  Lande $g$ factor, and $\mu_{\rm B}$ is the Bohr magneton.) Units are unimportant within our linear-response formalism, as the prefactors cancel out once we normalize the results via the recipe of Sec.~\ref{sec:penalty}.}
\begin{equation}
	{\bf m}({\bf r}) = \int [d^3 k] \sum_m f_{m\bf k} \psi^\dagger_{m\bf k}({\bf r}) \bm{\sigma} \psi_{m\bf k}({\bf r}),
	\label{mofr}
\end{equation}
where $\bm{\sigma}$ are the Pauli matrices and $\psi_{m\bf k}$ is a two-component spinor wave function for the electronic band $m$ at point ${\bf k}$ of the Brillouin zone.
The local magnetic moments are then defined as integrals of ${\bf m}({\bf r})$ over the Wigner-Seitz (WZ) atomic spheres, following the usual conventions,
\begin{equation}
{\bf m}_{\kappa} = \int_{\Omega(\kappa)} d^3 r \, {\bf m}({\bf r}) =  \int d^3 r \, {\bf m}({\bf r}) f_\kappa({\bf r}-\bm{\tau}_\kappa).
\label{eq:mmom}
\end{equation}
Here, $\kappa$ is an atomic basis index and $\bm{\tau}_\kappa$ the position
vector. 
The integration domain $\Omega(\kappa)$ is defined by $|{\bf r} -\bm{\tau}_\kappa|<R^{\rm WS}_\kappa$, where the $R^{\rm WS}_\kappa$ is the radius of the WS spheres.
This can be conveniently implemented by defining a window function $f_\kappa({\bf r})=\Theta(R^{\rm WS}_\kappa - |{\bf r}|)$, where $\Theta(x)$ is the Heaviside function.
[In practice, we set a smooth boundary, which is preferable in a plane-wave context.]

We then introduce the following modified Kohn-Sham functional,

\begin{equation}
\tilde{U}(\psi,\lambda,B_l) = E_{\rm KS}(\psi,\lambda) + \frac{\alpha}{2} \sum_l [B_l-m_l(\psi,\lambda)]^2,
\label{Utilde}
\end{equation}
where we have introduced a new set of free parameters, $B_l$, with the physical dimension of a magnetic moment. $l$ is a composite index that runs over the magnetic degrees of freedom, defined by their atomic sites and Cartesian components, and $\alpha$ is a constant to be adjusted for optimal performance on a case-by-case basis. 
Meanwhile, both the Kohn-Sham total energy ($E_{\rm KS}$) and the local magnetic moments ($m_l$) depend on the electronic orbitals ($\psi$) and on other physical parameters (atomic coordinates, electric fields, etc...) of the system (indicated as $\lambda$).
We shall be primarily interested on the related functional $\tilde{U}(\lambda,B_l)$, which we define as the variational miminum of $\tilde{U}(\psi,\lambda,B_l)$ with respect to the electronic degrees of freedom, under the usual orthonormality constraints,
\begin{equation}
\tilde{U}(\lambda,B_l) = \min_\psi \tilde{U}(\psi,\lambda,B_l).
\end{equation}
The purpose of the additional penalty term in Eq.~\eqref{Utilde} is to force the noncollinear spin system into a predetermined target configuration, which is specified by the parameters $B_l$. 
In practice, at the variational minimum of Eq.~\eqref{Utilde} $m_l$ and $B_l$ generally differ slightly, by an amount that becomes smaller the larger the constant $\alpha$.

The ground-state solution of the constrained functional allows one to explore the $2N_{\rm mag}$-dimensional parameter space of the possible magnetic configurations, where $N_{\rm mag}$ is the number of magnetic sites. 
This method, first established by Ma and Dudarev~\cite{MaPRB15}, is widely popular nowadays, and a useful feature of most first-principles codes.
(Note that standard implementations thereof target the direction, and not the modulus, of the local spin moment; this technicality is largely irrelevant for what follows so we will disregard it for simplicity.)
The reason behind our use of the symbol $B_l$ for the target magnetic moments rests on the analogy with the constrained-${\bf D}$ approach~\cite{fixedd} which can be regarded as the electrical counterpart of Eq.~\eqref{Utilde}.
Indeed, the constrained-${\bf D}$ functional is obtained by incorporating the Maxwell energy of the uniform electric field into $E_{\rm KS}$, and the latter [in the same form as the magnetic penalty on the rhs of Eq.~\eqref{Utilde}] effectively forces the macroscopic polarization (${\bf P}$) of the crystal to adopt a value as close as possible to the ``target polarization'', ${\bf D}$.  
The present interpretation of $B_l$ as a magnetic induction field is, of course, a loose one: for example, $m_l$ is a magnetic moment and not a magnetization; also, 
the prefactor $\alpha$ is not necessarily related to the vacuum permeability. 
Notwithstanding, we will often refer 
to the penalty-based internal energy $\tilde{U}$ as ``constrained-{\bf B} functional''.

Bringing the
analogy with the electric case~\cite{fixedd} one notch further, we can put 
our discussion of Eq.~\eqref{Utilde} on firmer theoretical grounds by
invoking the concept of \emph{Legendre transform}. 
In particular, starting from Eq.~\eqref{Utilde} we can define 
a magnetic enthalpy functional as the Legendre transform of $\tilde{U}$,
\begin{equation}
\begin{split}
\tilde{F}(\lambda,H_l) = &
\min_{B_l} \left[\tilde{U}(\lambda,B_l) - \sum_l H_l B_l\right] \\
=&\min_{\psi} \left[ E_{\rm KS}
- \sum_l H_l m_l \right] - \sum_l \frac{H_l^2}{2\alpha}.
\end{split}
\label{Ftilde}
\end{equation}
(We no longer indicate the parametric dependence of $E_{\rm KS}$ and $m_l$ on $\lambda$ and $\psi$ explicitly, as it is now obvious from the context.)
The new free parameters $H_l$ have the physical meaning of local Zeeman field amplitudes, and correspond to the first derivative of the internal energy
with respect to $B_l$ (see the next subsection).
The functional $\tilde{F}$ can be regarded as the magnetic counterpart of the electric field enthalpy~\cite{fixedd}. The quadratic term in $H_l$ plays the same role as
the macroscopic Maxwell energy in Ref.~\onlinecite{fixedd}, and has no impact on the electronic ground state since $H_l$ now enters the functional as an external parameter.

As an alternative to the penalty-function approach described above, a given magnetic configuration can also 
be enforced via holonomic constraints.
To that end, we first introduce the following enthalpy functional,
\begin{equation}
\label{eq:F}
{F}(\lambda,H_l) = \min_\psi \left[ E_{\rm KS} - \sum_l H_l m_l \right],
\end{equation}
which is
the same as $\tilde{F}(H_l)$, except for the trivial constant that is quadratic in the Zeeman field amplitudes.
Eq.~\eqref{eq:F} corresponds to the magnetic enthalpy for 
a system under a (local, in this case) Zeeman magnetic field
as defined in Ref.~\onlinecite{BousquetPRL11}.
Then, we define the internal energy $U$ as the inverse Legendre transform of $F$,
\begin{equation}
\label{eq:U}
\begin{split}
U(\lambda,M_l) =& \max_{H_l} \left[ F(\lambda,H_l) + \sum_l H_l M_l \right] \\
 =& \max_{H_l} \min_\psi  \left[ E_{\rm KS} + \sum_l H_l(M_l-m_l) \right].
\end{split}
\end{equation}
Here, $M_l$ are the free parameters, corresponding to the target set of
local magnetic moments, and the Zeeman fields $H_l$ play now the role of Lagrange multipliers,
whose purpose is to impose $m_l = M_l$ exactly. (The latter equality is guaranteed by the stationary condition 
on $H_l$.) This differs from the penalty-based approach,  Eq.~\eqref{Utilde}, where $m_l - M_l = H_l/\alpha$ can be made to be small, but never to vanish.
Eq.~\ref{eq:U} can be regarded as the magnetic counterpart of the constrained-${\bf P}$ functional proposed in Ref.~\onlinecite{dieguez/vanderbilt:2006}.
Recent developments in the framework of constrained-DFT~\cite{GonzeJCTC22} have renewed the interest for this class of approaches,
making them even more attractive for practical applications.

\subsection{First derivatives}

In order to perform a number of basic tasks (e.g., calculate the electronic ground state, or relax the atomic structure), it is essential to determine the first derivatives of the functionals introduced above. 
By taking the functional derivative of Eq.~\eqref{Utilde} with respect to the wave functions, one readily obtains the electronic Hamiltonian,
\begin{equation}
\label{hamil}
\hat{\mathcal{H}} = \hat{\mathcal{H}}_{\rm KS} + \sum_l H_l \hat{V}^{H_l},
\end{equation}
where we have introduced the calligraphic symbol, $\hat{\mathcal{H}}$, for screened Hamiltonians inclusive of the self-consistent (SCF) Hartree and exchange-correlation potentials. The additional local Zeeman potential reads as
\begin{equation}
V^{H_l}({\bf r}) = - \sigma_\beta
f_{\kappa}({\bf r}-\bm{\tau}_\kappa),
\label{eq:v0zeem}
\end{equation}
and $\sigma_\beta$ is a Pauli matrix. The Hamiltonian takes this same form in all four magnetic functionals introduced in this Section; the only difference consists in the way the Zeeman field strength $H_l$ enters the formalism.

In both enthalpies ($F$ and $\tilde{F}$), $H_l$ are external parameters that remain fixed throughout the simulation.
Conversely, in the context of the internal energies ($U$ and $\tilde{U}$), $H_l$ are themselves function of both the electronic ($\psi$) and structural ($\lambda$) degrees of freedom, and need to be determined self consistently. 
More generally, the first derivatives with respect to the magnetic variables
are in each case given by the standard Legendre transform relations:
\begin{equation}
\label{legendre1}
\begin{split}
\frac{\partial \tilde{U}}{\partial B_l} = H_l, \qquad & \frac{\partial \tilde{F}}{\partial H_l} = -B_l, \\
\frac{\partial {U}}{\partial M_l} = H_l, \qquad & \frac{\partial {F}}{\partial H_l} = -M_l.
\end{split}
\end{equation}
It is useful to work out the first derivative of $\tilde{U}$ explicitly,
where the the conjugate variable of $B_l$ is
\begin{equation}
\label{eq:hl}
 H_l=\frac{\partial \tilde{U}}{\partial B_l}=\alpha (B_l-m_l),
\end{equation}
Here the local Zeeman field consists of two separate contributions,
\begin{equation}
\label{hext}
H_l = H^{\rm ext}_l + H^{\rm ind}_l,
\end{equation}
where the ``external field'' (ext) is determined by the free parameter $B_l$
via $H^{\rm ext}_l = \alpha B_l$, while the ``induced'' (ind) part depends on 
the local magnetic moments via $H^{\rm ind}_l = -\alpha m_l$. 
This observation is central to our method; we shall come back to it shortly.

The derivatives with respect to other parameters $\lambda$ are given, via the Hellmann-Feynman theorem, in terms of the first derivative of the Hamiltonian, Eq.~(\ref{hamil}),
\begin{equation}
\label{eq:H1}
\frac{\partial \hat{\mathcal{H}}}{\partial \lambda}
 = \hat{H}_{\rm KS}^\lambda
 + \sum_l H_l \frac{\partial \hat{V}^{H_l}}{\partial \lambda}.
\end{equation}
The second term on the rhs appears whenever the local Zeeman fields are nonzero and the functions 
$f_\kappa({\bf r}-\bm{\tau}_{\kappa})$ explicitly depend on $\lambda$.
For example, if $\lambda$ refers to an atomic displacement, the additional term can be regarded as a Pulay force, originating from the fact that the magnetic species drag their own WS spheres along the trajectory.
Note that the partial derivative only acts on the explicit dependence of the Hamiltonian on the parameter $\lambda$, and excludes SCF fields. 

Similarly to the ground-state Hamiltonian, the first-order Hamiltonian has the same identical form in all cases, which means that all the magnetic functionals yield the same atomic forces, provided that the parameters used in each case are equivalent. 
In particular, suppose that we perform a calculation by using the standard magnetic enthalpy $F$, with a given set of Zeeman fields $H_l$. Once we reach the electronic ground state, we extract the converged values of the local magnetic moments, $m_l^0$. 
Then, we repeat the calculation, this time by using the internal energy $U$, by setting the corresponding magnetic input parameters to $M_l=m^0_l$; or we can alternatively use the modified enthalpy $\tilde{F}$ and the same set of $H_l$, or even adopt the constrained-${\bf B}$ functional, with $B_l = H_l / \alpha + m^0_l$: we will obtain exactly the same Hamiltonian, ground-state orbitals and atomic forces in all cases.

This is an important result, whose consequences are of key relevance for the remainder of this work. The discrepancy between the ``target magnetic moments'' ($B_l$) and their actual values ($m_l$) at the variational minimum of $\tilde{U}$, often regarded as the main drawback of the penalty-function approach~\cite{GonzeJCTC22}, can be easily dealt with via a set of exact Legendre relations. In other words, one is free to work with either flavor ($U$ or $\tilde{U}$) of the constrained-spin internal energy and rest assured that the physical results extracted from one or the other approach are in all respects equivalent. We shall come back to this point in Sec.~\ref{sec:choicefun}.

\subsection{Second derivatives \label{sec:susflavors}}

In the next few paragraphs, we shall establish exact relations between the second derivatives of the magnetic energy functionals presented insofar.
This is an essential ingredient of our method, as it allows one to perform calculations with the most computationally convenient functional, 
and subsequently reconstruct the physically meaningful properties of the system via a sequence of simple linear-algebra operations.
We organize the discussion into the three possible combinations of perturbations, either describing a purely magnetic response or involving other nonmagnetic degrees of freedom.
In the remainder of this Section, we assume that the derivatives are taken statically along an adiabatic path that preserves the periodicity of the crystal; the generalization to finite frequency $\w$ and momentum ${\bf q}$ is provided in the forthcoming Section~\ref{sec:causality}.

\subsubsection{Two magnetic fields}

We start by considering the second derivatives with respect to  two magnetic parameters.
In this context, the central quantity to be considered is the \emph{physical} susceptibility matrix. 
Its $jl$ element is defined as the first derivative of the local magnetic moment $j$ with respect to a Zeeman field $l$, 
\begin{equation}
\label{chi1}
	\chi_{jl} =  
	\frac{\partial m_j}{\partial H_l}\Big|_{F,\tilde F},
\end{equation}
where differentiation is meant to be carried out in the framework of the magnetic enthalpy functionals ($F$ or $\tilde F$).
By using Eqs.~(\ref{legendre1}), we can equivalently express $\chi_{jl}$ as (minus) the second derivative of the standard magnetic enthalpy $F$ with respect to two Zeeman fields,
\begin{equation}
\label{chi2}
	\chi_{jl} = -\frac{\partial^2 {F}}{\partial H_j \partial H_l} = -{F}_{jl}.
\end{equation}
(From now on, we shall use the shorthand notation $E_{jl}$ to indicate the second derivative of 
a given magnetic functional $E$ with respect to the independent magnetic variables.)
The second derivatives of the modified enthalpy, $\tilde{F}$, are also trivially related to $\bm{\chi}$ via
\begin{equation}
\label{ftil_chi}
\frac{\partial^2 \tilde{F}}{\partial H_j \partial H_l} = \tilde{F}_{jl} = -\chi_{jl} - \frac{\delta_{jl}}{\alpha}.
\end{equation}

For reasons that shall become clear shortly, it is numerically more advantageous to work with the internal energies, rather than with the enthalpies. As we anticipated earlier, the second derivatives of $U$ contain the same physical information as $F_{jl}$ or $\tilde{F}_{jl}$; for example, the following relation to the susceptibility holds,
\begin{equation}
\label{ujl}
\frac{\partial^2 {U}}{\partial M_j \partial M_l} = {U}_{jl} = \chi^{-1}_{jl}. 
\end{equation}
That $U^{-1}_{jl}=-{F}_{jl}$ is a general property of two functionals that are related via Legendre transformation, and holds for the tilded pair as well,
\begin{equation}
\label{ujl_til}
\frac{\partial^2 \tilde{U}}{\partial B_j \partial B_l} = \tilde{U}_{jl} = -\tilde{F}_{jl}^{-1}.  
\end{equation}
By combining the latter expression with Eq.~\eqref{ftil_chi}, we obtain a Dyson-like expression for the susceptibility in terms of $ \tilde{U}_{jl}$,
\begin{align}
\label{dyson_chi}
	\bm{\chi} =& \tilde{\bf U}^{-1} - \frac{1}{\alpha} {\bf I}.
\end{align}

For what follows, it is also useful to define an auxiliary quantity that is proportional to the inverse of Eq.~(\ref{utilu}),
\begin{equation}
\label{chi_til}
\tilde{\bm{\chi}} = \left( \bm{\chi}^{-1} + \alpha {\bf I} \right)^{-1},
\end{equation}
and has the physical interpretation of a spin-spin correlation function within the
constrained-{\bf B} functional $\tilde{U}$.
Indeed, one can equivalently express $\tilde{\bm{\chi}}$ as derivative of the local magnetic moments $m_l$
with respect to the external Zeeman field as defined in Eq.~\eqref{hext},
\begin{equation}
\tilde{\chi}_{jl} = \frac{1}{\alpha} \frac{\partial m_j}{\partial B_l} \Big|_{\tilde{U}} =  \frac{\partial m_j}{\partial H^{\rm ext}_l} \Big|_{\tilde{U}}.
\end{equation}
After observing that
\begin{equation}
\label{ujl_til2}
\frac{\partial^2 \tilde{U}}{\partial B_j \partial B_l} = \alpha(\delta_{jl} - \alpha \tilde{\chi}_{jl} ),
\end{equation}
we can also summarize the above results more compactly as follows,
\begin{equation}
\label{summary_chi}
\tilde{\bf U}^{-1} = \frac{1}{\alpha}( {\bf I} + \alpha \bm{\chi}) = \frac{1}{\alpha}( {\bf I} - \alpha \tilde{\bm{\chi}})^{-1}.
\end{equation}

\subsubsection{Mixed derivatives}

We next consider the case of second derivatives with respect 
to one magnetic-like perturbation and another external perturbation, $\lambda$, 
referring to atomic basis and/or spatial directions.
These derivatives can be directly related to the
first-order local magnetic moments induced by the perturbation $\lambda$ at specific boundary conditions.
For instance, the relaxed-spin first-order magnetic moments
are given by
\begin{equation}
 \frac{\partial m_j}{\partial \lambda}\Big|_{F}=-\frac{\partial^2 {F}}{\partial H_j \partial \lambda}=-{F}_{j\lambda},
 \label{eq:shorthand}
\end{equation}
where the $F$ subscript on the left hand side indicates that
the derivative is intended to be taken within the framework of the 
standard magnetic enthalpy.
This implies that the first-order magnetic moments are written as
minus the second derivative of the enthalpy. 
Note that ${F}_{l\lambda}=\tilde{F}_{l\lambda}$: we only need to distinguish between the two enthalpies
when both indices are magnetic in nature.

As in the case of the purely magnetic 
response functions, the same information can be extracted from the 
second derivatives of the internal energy.
For the constrained-${\bf M}$ functional, we have
\begin{align}
& \frac{\partial m_j}{\partial \lambda}\Big|_{F}=-
\chi_{jl} \frac{\partial H_l}{\partial \lambda}\Big|_{U} = -U^{-1}_{jl} U_{l \lambda}, \label{mlam1}
\end{align}
Here, $\frac{\partial H_p}{\partial \lambda_a}\Big|_{U}$ have the physical interpretation of internal local Zeeman fields induced by the perturbation $\lambda$ when the magnetic moments are fixed; this is the type of data usually extracted from finite-difference methods such as the one used by Ren et al.~\cite{RenPRX24}.
In fact, a relation that is in all respects equivalent to Eq.~\eqref{mlam1} also holds for the constrained-${\bf B}$ functional,
\begin{align}
& \frac{\partial m_j}{\partial \lambda}\Big|_{F}=-\tilde{U}^{-1}_{jl}
\frac{\partial H_l}{\partial \lambda}\Big|_{\tilde U} = ( {\bf I} - \alpha \tilde{\bm{\chi}})^{-1}_{jl} \frac{\partial m_l}{\partial \lambda}\Big|_{\tilde U},
\label{mlam2}
\end{align}
where the second equality follows immediately from Eq.~(\ref{summary_chi}).
Finally, by combining Eq.~(\ref{mlam1}) and Eq.~(\ref{mlam2}), we obtain a formula that directly yields the first-order Zeeman fields at frozen spins in terms of the constrained-${\bf B}$ (penalized-spin) response, 
\begin{align}
\label{conv_mixed}
& \frac{\partial H_j}{\partial \lambda}\Big|_{U}= U_{jl} \tilde{U}^{-1}_{lp} \frac{\partial H_p}{\partial \lambda}\Big|_{\tilde U}.
\end{align}
The matrix product on the rhs can be further simplified by using Eq.~\eqref{dyson_chi}, and written in a manifestly symmetric form,
\begin{equation}
\label{utilu}
U_{jl} \tilde{U}^{-1}_{lp} = \delta_{jp} + \frac{1}{\alpha} \chi^{-1}_{jp} = (\alpha \tilde{\bm{\chi}})_{lp}^{-1}.
\end{equation}
This expression is consistent with the expected behavior in the large-$\alpha$ limit:
the first-order magnetic moments in response to $\lambda$ become increasingly suppressed
by the penalty, which implies that the first-order Zeeman fields within the contrained-{\bf B} approach recover 
the same values as in the frozen-spin functional. The second equality shows that $\tilde{\chi}_{jl}\simeq \delta_{jl} / \alpha$
in the same limit.
 
\subsubsection{Nonmagnetic second derivatives}

Regarding the second derivatives with respect to two nonmagnetic 
perturbations,  
the relevant relations are 
\begin{subequations}
\begin{align}
& {F}_{\lambda \lambda'}={U}_{\lambda \lambda'}- U_{\lambda l} 
 {\chi}_{lp} U_{p \lambda'}
 \label{eq:E2_BtoH_a}\\
& {U}_{\lambda \lambda'}= \tilde{U}_{\lambda \lambda'} + U_{\lambda l}  
 \tilde{\chi}_{lp} 
 U_{p \lambda'}
\label{eq:E2_BtoH_b}
\end{align}
\label{eq:E2_BtoH}
\end{subequations}
where the mixed derivatives are given by
\begin{equation}
U_{j\lambda} = \frac{\partial H_j}{\partial \lambda}\Big|_{U}.
\end{equation}

The above equations constitute one of the main results of the present work: In combination with Eq.~\eqref{conv_mixed}, they allow one to recover the physically meaningful response functions (dielectric tensor, force constants, Born effective charges, etc.) from those calculated at constrained-${\bf B}$.
In addition, Eq.~\eqref{eq:E2_BtoH_a} represents a physically interesting partition of ${F}_{\lambda \lambda'}$. 
${U}_{\lambda \lambda'}$ corresponds to the frozen-spin (FS) response, i.e., calculated by enforcing the magnetic moments to be the exactly the same as in the unperturbed state; for this reason, this term is free from magnonic resonances.
In turn, the second term on the right hand side describes the additional (possibly resonant) contributions due to the induced spin cantings; we shall consequently refer to the total ${F}_{\lambda \lambda'}$ as the relaxed-spin (RS) response coefficient.

Note that ${U}_{\lambda \lambda'}$ is generally sensitive to the details (e.g., the form factor of the atomic spheres) of how the magnetic moments are constrained to their unperturbed value. Conversely, ${F}_{\lambda \lambda'}$ is independent of such choices, and must yield the exact same outcome (bar the convergence issues discussed below) of a straight DFPT calculation~\cite{RicciPRB19} of the second-order Kohn-Sham energy.

\subsection{Choice of the functional \label{sec:choicefun}}

In the above sections we have demonstrated that all magnetic functionals are formally equivalent, as the ground-state properties and response functions that one extracts from one or the other are related by simple linear-algebra expressions. 
From the point of view of a practical numerical implementation, however, it is highly advantageous to work with the internal energies rather than with the enthalpies. 
In most materials, the spin degrees of freedom are characterized by a very shallow energy landscape with multiple local minima. This means that the electronic Hessian of the unconstrained Kohn-Sham functional $F$ is poorly conditioned, and reaching convergence can be difficult or impossible depending on the system. 
The situation becomes all the more serious in the dynamical regime, as the frequency-dependent susceptibility diverges at the physical magnon frequencies. 

Both internal energy functionals described here provide an effective solution to the aforementioned issue.
In the case of $U$, the reason is obvious: the holonomic constraints on the local magnetic moments freeze the spin degrees of freedom, effectively removing the problematic low-lying modes from the electronic Hamiltonian. Furthermore, the ${U}_{jl}$ matrix corresponds to the inverse of $\bm{\chi}$ and is a smooth function of frequency within the sub-gap range of the spectrum.
Within the penalty-based internal energy, $\tilde{U}$, the magnetic moments are not frozen completely, and the performance of the method relies on choosing an appropriate value of the parameter $\alpha$. 
The main indicator in this context is given by the spin-spin correlation function introduced in Eq.~\eqref{chi_til}. In the limit of small $\alpha$, $\tilde{\bm{\chi}}$ coincides with the physical spin-spin susceptibility of the system, as the additional penalty term in Eq.~\eqref{Utilde} becomes vanishingly small (e.g., compared to the magnetocrystalline anisotropy). For increasing $\alpha$, the potential energy landscape associated with the magnetic degrees of freedom progressively stiffens; as a consequence, the poles of $\tilde{\bm{\chi}}$ shift to higher energy. If $\alpha$ is large enough, such resonances become larger than the optical gap, and therefore are no longer an issue for the numerical calculations. (Pushing $\alpha$ to even higher values does not improve the condition number of the problem, and might lead to numerical instabilities.)

Since both $U$ and $\tilde{U}$ yield the same physical answers, the choice of one over the other is a matter of convenience. In this work we have opted for the penalty-based approach for our calculations, as we found it easier to implement within the linear-response module of ABINIT in its current state. Indeed, no modification is needed to the preexisting potential mixing scheme, unlike the Lagrange multiplier method described in Ref.~\cite{GonzeJCTC22}. Notwithstanding, we find it most appropriate to convert all physical properties of relevance into second derivatives of $U$ prior to storing them for later use. This way we avoid the undesirable dependence of $\tilde{U}$ on the arbitrary constant $\alpha$, and we obtain a more direct link to recent literature works~\cite{GonzeJCTC22,RenPRX24}.

\subsection{Dynamical regime and causality\label{sec:causality}}

To access a number of useful physical properties, it is necessary in general to assume perturbations of the system 
that are modulated in space and time as follows,
\begin{equation}
\label{deltav}
\Delta V({\bf r},t) = \lambda^{{\bf q},\w} V^\lambda_{{\bf q}\w}({\bf r}) e^{i{\bf q \cdot r}- i(\omega + i\eta^+) t}.
\end{equation}
Here $\Delta V({\bf r},t)$ is the perturbing potential entering the crystal Hamiltonian, which is conveniently 
expressed as a cell-periodic part $V^\lambda_{{\bf q}\w}$ times a complex phase. The latter depends on 
the momentum ${\bf q}$ and on the frequency $\w$; the positive infinitesimal $\eta^+$ ensures proper 
treatment of causality; $\lambda^{{\bf q},\w}$ is the (generally complex) perturbation parameter. 

Spatial modulation (embodied by the $e^{i {\bf q \cdot r}}$ phase factor) is easy to treat within the context of 
density-functional perturbation theory (see Sec.~\ref{sec:implement}) and entails only minor modifications to the 
formulas of Sec.~\ref{sec:susflavors}.
The assumption of a finite frequency, on the other hand, implies recasting the formalism of the 
earlier sections into the language of time-dependent perturbation theory, which requires a number of
additional considerations. We shall briefly go through them in the following.

For an insulator in the transparent regime (i.e., at frequencies that are smaller than 
the lowest cross-gap resonances) the electronic response to an external perturbation 
is purely reactive; this means that we can safely set $\eta^+$ to zero in Eq.~\eqref{deltav}.
(In presence of noncollinear spins, the low-lying magnon resonances generally violate such a condition;
as we explained in the previous Section, however, this issue is solved once and for all by working at constrained 
local magnetic moments.) 
Then, one can conveniently express the frequency- and momentum-dependent response functions
as second derivatives of the Kohn-Sham \emph{action} functional, which is defined as~\cite{DalCorsoPRB96}
\begin{equation}
\label{action}
\begin{split}
A_{\rm KS}(\lambda,\psi) = \frac{1}{T} \int_0^T dt \Big[ & -i \int [d^3k] \sum_m \langle \psi_{n\bf k}(t) | \dot{\psi}_{n\bf k}(t) \rangle \\
 & + E_{\rm KS}(t) - \sum_{\kappa \alpha} \frac{M_\kappa}{2} \dot{\tau}^2_{\kappa \alpha} \Big],
\end{split}
\end{equation}
with $M_\kappa$ and $\tau_{\kappa \alpha}$ being respectively the mass and displacements along $\alpha$ of sublattice $\kappa$.
At this point, in order to generalize the theory of Sec.\ref{sec:susflavors} to the dynamical case, it suffices to 
replace $E_{\rm KS}$ with $A_{\rm KS}$ in the definition of all magnetic functionals. (We shall keep referring
to these functionals as internal energies and enthalpies, even if such a nomenclature 
is no longer accurate at finite frequency.)
The main practical consequence of working at finite $\w$ and/or ${\bf q}$ is that the 
response coefficients are now complex quantities, defined by
\begin{equation}
U_{ab}(\w,{\bf q}) = \frac{\partial^2 {U}}{\partial \lambda_a^{\bf -q,-\w} \partial \lambda_b^{\bf q,\w}}. 
\end{equation}
Given the absence of dissipation, the second-order matrices are Hermitian tensors,
\begin{equation}
\label{herm}
U^*_{ab}(\w,{\bf q}) = U_{ba}(\w,{\bf q})
\end{equation}
By keeping this in mind, most of the formulas presented in Sec.\ref{sec:susflavors} can be directly 
applied to the dynamical response at finite momentum without modification.

A notable exception concerns the Legendre relations that are used to switch between the
frozen-spin internal energies and the relaxed-spin enthalpies. These typically involve the 
inverse of the spin-spin correlation matrix, which becomes singular at the magnon resonances.
A similar issue occurs in the inversion of the frequency-dependent  interatomic force-constants (IFC) matrix, which we
shall use in the next Section to define the phonon propagator.
To perform such operations, a careful treatment of causality at the level of the $U_{ab}$ coefficients
is important; this can be easily done a posteriori, via analytic continuation to the upper complex half-plane
in frequency space.

In particular, 
we first
represent the frozen-spin internal energy as a Taylor expansion in powers of the frequency $\w$,
\begin{equation}
\label{u_causal}
	U_{ab}(\w) = \sum_{m=0}^\infty \frac{\w^m}{m!} U_{ab}^{(m)} .
\end{equation}
Once the coefficients of the Taylor expansion are determined via a numerical fit, then we can simply replace $\w \rightarrow \w + i\eta$ in the 
right-hand side (with $\eta$ a suitably small number) and thereby obtain the desired causal functions, whose matrix inverse is well defined at all frequencies. 
Note that this procedure always yields \emph{retarded} correlation matrices, where the second index [$b$ in Eq.~\eqref{u_causal}] is assumed to be the cause, i.e., an external field that we adiabatically switch on at a time $t=-\infty$, and the first index ($a$) refers to the effect (response).
For the same reason, Eq.~\eqref{herm} no longer holds if $\eta\neq 0$, as the response functions acquire an anti-Hermitian dissipative part.

\subsection{Spin-phonon correlation functions\label{sec:corrfun}}

In the following, we 
demonstrate that the formalism developed in the earlier sections can be used to determine the frequency-dependent correlation functions between the active degrees of freedom of the crystal (phonons and spins). 
As we shall see shortly, such knowledge is essential in order to calculate the
susceptibility of the system with respect to an arbitrary external perturbation and in a variety of conditions 
(fixed- or relaxed-spin, clamped- or relaxed-ion or any combination thereof). 
The first task consists in generalizing the formalism developed insofar to allow for the simultaneous treatment of the atomic displacements $\tau_{\kappa \alpha}$ alongside with the local magnetic moments $M_j$.

In close analogy with the formalism that we established in Section~\ref{sec:gsfun}, we frame our theory around the concept of Legendre transform, which we
apply to the extended set of variables $(\tau,M)$.
By construction, the functional $U(\tau,M)$ of Section~\ref{sec:gsfun}  plays the role of the internal energy in both the spin and phonon sectors.
Its Legendre transform with respect to the lattice variables reads as 
\begin{equation}
\label{gen_enthalpy}
\mathcal{U}(f, M) = \min_\tau \left[ U(\tau,M) - \sum_{\kappa \alpha} f_{\kappa \alpha} \tau_{\kappa \alpha} \right].
\end{equation}
The conjugate variable to $\tau_{\kappa \alpha}$ is $f_{\kappa \alpha}$, and plays the role of an external force along $\alpha$ acting on the sublattice $\kappa$.
We can perform a similar operation on the magnetic enthalpy $F$ and obtain the \emph{spin-phonon enthalpy}, 
\begin{equation}
\label{fcal}
\mathcal{F}(f, H) = \min_\tau \left[ F(\tau,H) - \sum_{\kappa \alpha} f_{\kappa \alpha} \tau_{\kappa \alpha} \right].
\end{equation}
The mutual relations between the four functionals defined insofar is in all respects analogous to the textbook thermodynamic relations between internal energy, enthalpy, Helmholtz and Gibbs free energy.

The physical significance of the four functionals defined above becomes clear once we look at their second derivatives.
In all cases, the resulting matrices take a four-block structure; for example, the Hessian of ${\bf U}(\tau,M)$ can be written as, 
\begin{equation}
\label{Uspinph}
 {\bf U} =
 \begin{pmatrix}
 	{\bf U}^{(ss)} & {\bf U}^{(sp)} \\
 	{\bf U}^{(ps)} & {\bf U}^{(pp)}
 \end{pmatrix},	
\end{equation}
where the individual elements refer to the phonon-phonon ($pp$), spin-spin ($ss$) and mixed second derivatives ($ps$ and $sp$) introduced in Section~\ref{sec:susflavors}.
Note that the $pp$ channel takes the form
\begin{equation}
\label{uphim}
U^{(pp)}_{\kappa \alpha, \kappa' \beta}({\bf q},\omega) = {\Phi}^{\rm FS}_{\kappa \alpha, \kappa' \beta}({\bf q},\omega)- M_\kappa \delta_{\kappa\kappa'} \delta_{\alpha\beta} \w^2,
\end{equation}
here, the first term on the rhs is the frozen-spin IFC matrix, and the mass contribution originates from the last term in Eq.~\eqref{action}.
Similarly, we define as ${\bf F}$ the Hessian of the functional $F(\tau,H)$, which is related to $U(\tau,M)$ via a partial Legendre transform operated on the spin channel only;
its four blocks are related to those of ${\bf U}$ via the relations established in Section~\ref{sec:susflavors}.
In particular, its $pp$ channel reads as
\begin{equation}
\label{fphim}
F^{(pp)}_{\kappa \alpha, \kappa' \beta}({\bf q},\omega) = {\Phi}^{\rm RS}_{\kappa \alpha, \kappa' \beta}({\bf q},\omega)- M_\kappa \delta_{\kappa\kappa'} \delta_{\alpha\beta} \w^2,
\end{equation}
where the relaxed-spin IFC matrix is given by $\bm{\Phi}^{\rm RS} = \bm{\Phi}^{\rm FS} - {\bf U}^{(ps)} \bm{\chi} {\bf U}^{(sp)}$ and
$\bm{\chi}=-{\bf F}^{(ss)}=[{\bf U}^{(ss)}]^{-1}$ is the spin-spin correlation function at clamped ions.

Of the remaining two functionals $\mathcal{F}(f,H)$ is the most interesting physically. 
Its Hessian, $\bm{\mathcal{F}}$, corresponds to the full correlation function in the extended spin-phonon parameter space,  
defined as the amplitude of the atomic displacement (or spin canting) $q_a$ in response to an externally applied atomic force (or local Zeeman field) $f_b$, at a given frequency $\w$ and momentum ${\bf q}$,
\begin{equation}
\label{susc}
\frac{\partial q_a}{\partial f_b} = -\mathcal{F}_{ab} = U^{-1}_{ab}.
\end{equation}
Its poles correspond to the resonances (in general of mixed magnon-phonon character) of the coupled spin-lattice system, and can therefore be regarded as a Green's function within this extended parameter space. 
To touch base with the extensive electron-phonon literature, note that 
\begin{equation}
\label{fcal_pp}
\begin{split}
-\bm{\mathcal{F}}^{(pp)} =& \left[ {\bf U}^{(pp)} - {\bf U}^{(ps)} [{\bf U}^{(ss)}]^{-1} {\bf U}^{(sp)} \right]^{-1} = [{\bf F}^{(pp)}]^{-1}
\end{split}
\end{equation}
coincides (modulo a sign) with the nonadiabatic phonon propagator indicated as ${\bf D}(\w)$ in Ref.~\onlinecite{GiustinoRMP17}. 
It is also interesting to consider the spin-spin block,
\begin{equation}
\label{fcal_ss}
\begin{split}
-\bm{\mathcal{F}}^{(ss)} =& \left[ {\bf U}^{(ss)} - {\bf U}^{(sp)} [{\bf U}^{(pp)}]^{-1} {\bf U}^{(ps)} \right]^{-1} 
 = [\bm{\mathcal{U}}^{(ss)}]^{-1}.
\end{split} 
\end{equation}
which corresponds to the spin-spin correlation function (or local spin susceptibility) at relaxed ions, $\bm{\chi}^{\rm RI}$.

All in all, the present formalism provides a simple way to access all possible combinations of clamping or relaxing a subset of modes 
via a unified procedure, which is of central importance when discussing the dynamical response of the system to an external field.

\subsection{Dielectric, magnetic \blue{and magnetoelectric} susceptibilities \label{sec:sus}}

We shall now consider the susceptibility of the system with respect to two types of macroscopic perturbations, respectively a uniform Zeeman, ${\bf H}^{\rm mac}$, and a uniform electric field, $\bm{\mathcal{E}}$. The coupling can be incorporated by following the same strategy, based on Legendre transforms, as above; the guidelines in the electrical case have been discussed at length in Ref.~\onlinecite{fixedd}.
In the context of the present work, we shall limit ourselves to adding the following terms to the Kohn-Sham energy $E_{\rm KS}$~\cite{SouzaPRL02,BousquetPRL11},
\begin{equation}
\label{deltae}
\begin{split}
\Delta E(\psi,\tau,\mathcal{E},H^{\rm mac}) = -\Omega {\bf P}(\psi,\tau)\cdot \bm{\mathcal{E}} - {\bf H}^{\rm mac} \cdot {\bf m}(\psi),
\end{split}
\end{equation}
where ${\bf P}$ is the formal polarization and ${\bf m}$ is the total magnetic moment of the cell.
This modification then propagates to all Legendre pairs defined so far either directly (in the static case) or via the quantum action, Eq.\eqref{action}~\footnote{
Strictly speaking, after incorporation of Eq.~\eqref{deltae}, $U$ is no longer an \emph{internal} energy (or action), as it allows for exchange of work between the system and external voltage (or magnetic field) sources.
Nevertheless, we shall keep referring to it with the same name and symbol, in order to avoid overburdening the notation.}.
At this stage, the parametric second derivatives of $U$, $F$, $\mathcal{U}$ or $\mathcal{F}$ with respect to $\bm{\mathcal{E}}$ or ${\bf H}^{\rm mac}$ directly yield the desired electric or magnetic susceptibilities within any physically relevant regime.

For instance, the dielectric susceptibility at clamped ions (CI) and frozen spins (FS) is readily given by
\begin{equation}
\bar{\chi}^{\rm die}_{\alpha \beta}(\w) = -\frac{1}{\Omega} U_{\mathcal{E}_\alpha \mathcal{E}_\beta}(\w).
\end{equation}
Conversely, the full susceptibility including the dynamical screening mediated by the coupled spin-lattice response (i.e., calculated within a RS and RI regime)
reads as
\begin{equation}
{\chi}^{\rm die}_{\alpha \beta}(\w) = -\frac{1}{\Omega} \mathcal{F}_{\mathcal{E}_\alpha \mathcal{E}_\beta}(\w).
\end{equation}
Based on the results derived above, it's easy to relate the two via 
\begin{equation}
\label{eq:chidie}
{\chi}^{\rm die}_{\alpha \beta} =  \bar{\chi}^{\rm die}_{\alpha \beta} - \frac{1}{\Omega} U_{\mathcal{E}_\alpha a} \mathcal{F}_{ab} U_{b \mathcal{E}_\beta}.
\end{equation}
The mixed derivative of $U$ in the second term on the rhs has the meaning of a generalized Born dynamical charge tensor, describing the macroscopic polarization that is associated with an atomic displacement or spin canting,
\begin{equation}
Z_{\alpha a}(\w) = \Omega \frac{\partial P_\alpha}{\partial \lambda_a} = -U_{\mathcal{E}_\alpha a}(\w).
\end{equation}
Here we have used the known thermodynamic relation between the electric field derivatives of the energy and the polarization~\cite{fixedd},
\begin{equation}
\frac{\partial U}{\partial \mathcal{E}_\alpha} = -\Omega P_\alpha.
\end{equation} 
Finally, we can plug Eq.~(\ref{eq:chidie}) into the standard definition of the frequency-dependent dielectric tensor,
\begin{equation}
\label{eq:epsdie}
\epsilon_{\alpha\beta}(\w) = \delta_{\alpha\beta} + 4\pi {\chi}^{\rm die}_{\alpha \beta} (\w).
\end{equation}
One can easily verify that Eq.~(\ref{eq:epsdie}) reduces to to the standard formulas~\cite{gonze/lee,maradudin} for the infrared-range dielectric tensor in absence of magnetic degrees of freedom.
Eq.~(\ref{eq:epsdie}) constitutes the generalization thereof to magnetic insulators with coupled spin-phonon excitations, and is one of the key formal results of this work.

\blue{The macroscopic magnetic susceptibility can be expressed in a similar form as Eq.~\eqref{eq:chidie},
\begin{align}
\label{chimag}
{\chi}^{\rm mag}_{\alpha \beta} =  \bar{\chi}^{\rm mag}_{\alpha \beta} - \frac{1}{\Omega} U_{{H}_\alpha a} \mathcal{F}_{ab} U_{b {H}_\beta},
\end{align}
where $\bar{\chi}^{\rm mag}_{\alpha \beta}(\w)= -U_{H_\alpha H_\beta}(\w)/\Omega$ describes the clamped-ion and frozen-spin response, 
and the subscript $H_\alpha$ indicates differentiation with respect to the macroscopic (uniform) Zeeman field.
Of the two additional ingredients involved in Eq.~\eqref{chimag}, note that $\bar{\chi}^{\rm mag}_{\alpha \beta}(\w)$ tends to be negligibly small,
as the dominant contribution to the response is typically associated with the canting of the local spin moments. For the same reason, the tensor
$U_{{H}_\alpha a}$ often adopts a trivial structure, and reduces to a proportionality constant between the macroscopic and
local Zeeman fields along the same Cartesian direction $\alpha$.
This means that the information about the macroscopic magnetic susceptibility can be directly extracted, to a high degree of accuracy,
from the spin-spin sector of the matrix $\mathcal{F}_{ab}$, consistent with the usual assumptions of spin-lattice models.
We shall provide direct numerical tests of the latter statement in the forthcoming application sections V and VI, respectively focusing on CrI$_3$ and Cr$_2$O$_3$.}

\blue{Finally, within our formalism the macroscopic magnetoelectric (ME) susceptibility reads as
\begin{equation}
\label{eq:me}
{\alpha}_{\alpha \beta} =  \bar{\alpha}_{\alpha \beta} - \frac{1}{\Omega} U_{{H}_\alpha a} \mathcal{F}_{ab} U_{b \mathcal{E}_\beta}.
\end{equation}
where $\bar{\alpha}_{\alpha \beta}(\w) = -U_{H_\alpha \mathcal{E}_\beta}(\w)/\Omega$ describes, as above, the clamped-ion and frozen-spin contributions. 
(The orbital magnetoelectric coupling may or may not be contained in $\bar{\alpha}_{\alpha \beta}$ depending on how the total magnetic moment of the 
cell is defined in Eq.~\eqref{deltae}.)
This way, Eq.~\eqref{eq:me} generalizes the standard descriptions of the magnetoelectric response (see, e.g.,  Refs.~\onlinecite{MengPRB14,WojdelPRL09})
by incorporating the dynamical response of the coupled phonon-magnon system.}

\subsection{Adiabatic expansion \label{sec:linearw}}

An appealing feature of the spin-phonon ${\bf U}$ matrix defined in Eq.~\eqref{Uspinph} consists in its regular character  
and weak dependence of frequency within the relevant range of the spectrum.
This means that ${\bf U}$ is ideally suited to an adiabatic expansion in powers of $\omega$,
\begin{equation}
{\bf U}(\w) = {\bf K} + i\w {\bf G} -  \w^2 {\bf M} + \cdots,
 \label{eq:linearw}
\end{equation}
where the dots stand for higher orders in $\w$.
Note that the mass operator ${\bf M}$ has the same basic structure as ${\bf K}$ and ${\bf G}$, i.e., it can be decomposed into spin-spin, phonon-phonon and off-diagonal subblocks,
\begin{equation}
 {\bf M} =
 \begin{pmatrix}
 	{\bf M}^{(ss)} & {\bf M}^{(sp)} \\
 	{\bf M}^{(ps)} & {\bf M}^{(pp)}
 \end{pmatrix}.
\end{equation}
All blocks are nonzero in general, since in addition to the (dominant) contribution of the bare nuclear masses
to ${\bf M}^{(pp)}$, they also contain an electronic contribution from the frequency dispersion of the $U_{ab}$
coefficients.

The convenience of Eq.~(\ref{eq:linearw}) lies in the hope that
higher orders in $\w$ can be dropped 
without compromising the accuracy of the result.
In the remainder of this work, 
we shall test two levels of approximation, which we 
refer to as first-order (FOA) and second-order (SOA) adiabatic theory.
Within FOA, we truncate Eq.~\eqref{eq:linearw} to first order in $\w$,
while discarding the electronic contributions to ${\bf M}$.
In other words, we set the ${\bf M}^{(pp)}$ tensor 
to the conventional diagonal matrix of atomic masses 
[corresponding to the second term on the rhs of Eq.~(\ref{uphim})], 
while the remainder blocks of ${\bf M}$ are set to zero. 
The advantage is that only Hessians and Berry curvatures (zero-th and first 
order in $\w$) need to be explicitly calculated, which can be done very efficiently 
with the methods described in Sec.~\ref{sec:implement}.
This approach corresponds to the spin-wave theory of Niu and Kleinman~\cite{NiuPRL98,QianPRL02,lin2025},
later generalized to the treatment of the coupled spin-phonon system by Ren \emph{et al.}~\cite{RenPRX24}

Within SOA, we follow the same procedure as in FOA, except that we fully account for
the electronic contributions to the mass tensor ${\bf M}$. 
The physical explanation of these additional terms lies in the inertia of the electrons that are dragged by a given degree of freedom along its trajectory. 
\blue{(In the phonon sector, the present theory recovers the adiabatic renormalization of the ionic masses as established in Ref.~\cite{Scherrer-17}.)}
Since the electrons are much lighter than the nuclei, this renormalization of the masses has, in most cases, a marginal impact on the phonon frequencies. Conversely, in the spin-spin channel the SOA constitutes a \emph{qualitative} departure from the usual assumption of zero magnon mass, and (as we shall demonstrate in Sections~\ref{sec:CrI3} and~\ref{sec:Cr2O3}) may have a significant impact on the calculated frequencies. At the SOA level, the time-honored Landau-Lifshitz (or Niu-Kleinman) equations of motion for the local spin moments $s_l$ must be revised as follows,
\begin{equation}
\sum_l \left( M_{jl} \ddot{s}_l - G_{jl} \dot{s}_l + K_{jl} s_l \right) = 0,
\end{equation}
where all matrices refer to their respective $(ss)$ subblock. We regard this as one of the major formal achievements of this work.
Our approach retains the conceptual simplicity of the FOA method established in Ref.~\onlinecite{RenPRX24}, in that the coupled magnon and phonon frequency spectrum can be calculated by solving essentially the same secular equation. On the other hand, the renormalization of the masses described here substantially improves the accuracy of the results to a level that, at least in our CrI$_3$ \blue{and Cr$_2$O$_3$ testcases}, is indistinguishable from the exact DFPT treatment.

\section{Implementation \label{sec:implement}}

In the following, we briefly recap the basics of density-functional perturbation theory (DFPT), which provides the general theoretical framework for carrying out the perturbative expansion of the Kohn-Sham energy.
We frame our presentation around the dynamical formulation of DFPT \blue{by assuming a plane-waves basis set and norm-conserving pseudopotentials}, which is appropriate to dealing with the low-energy magnon and phonon resonances. In other words, we approach the problem within the general framework of time-dependent density-functional theory at the adiabatic local-density approximation level, by working in the frequency domain and in the linear-response regime.

Our formalism is closely related to those employed previously to analyze the nonadiabatic linear response to atomic displacements~\cite{CalandraPRB10}, electric~\cite{DalCorsoPRB96,AndradeJCP07} and Zeeman magnetic~\cite{CaoPRB18,GorniEPJB18,LiuPRB23} fields.
We loosely follow the notation of Ref.~\onlinecite{RicciPRB19}; however, we omit for
simplicity the explicit indication of spin indices (both in the wave functions and 
$2\times2$ density matrices and operators), except for situations where it becomes essential.

\subsection{Dynamical DFPT}

To begin with, we represent the perturbing potential of Eq.~(\ref{deltav}) as a first-order Hamiltonian~\footnote{This is necessary to deal with 
nonlocal pseudopotential terms in the phonon response.},
(in the following, we shall alternate between representing the $\omega$-dependence using a subscript notation, when paired with explicit spatial coordinate dependence, and a parenthetical notation otherwise.)
\begin{equation}
\Delta \mathcal{H}({\bf r,r'}) = 
	\lambda^{\bf q,\w}
e^{i{\bf (k+q)\cdot r}} \, \mathcal{H}^\lambda_{\bf k,q\w}({\bf r,r'}) \, e^{-i{\bf k\cdot r'}} \, e^{- i\omega t},
\end{equation}
which couples states at any given point ${\bf k}$ in the Brillouin zone with those at ${\bf k+q}$.
The cell-periodic part of the operator satisfies the following Hermiticity condition, 
\begin{equation}
\label{tri}
\hat{\mathcal{H}}_{\bf k-q,q}^{\lambda} (\omega) = [\hat{\mathcal{H}}_{\bf k,-q}^{\lambda} (-\omega)]^\dagger.
\end{equation}
In turn, the screened first-order Hamiltonian ($\mathcal{H}^\lambda$) consists of an external potential ($H^\lambda$) that generally does not depend on frequency, plus a self-consistent (SCF) field contribution ($V^\lambda$) that is independent of ${\bf k}$,
\begin{equation}
	\mathcal{H}^\lambda_{\bf k,q}(\w) = {H}^\lambda_{\bf k,q} + V^\lambda_{\bf q}(\w).
\end{equation}
The latter is defined in terms of the first-order density $n^{\la}_{{\bf q}\omega}({\bf r})$ and the interaction (Hartree plus exchange and correlation) kernel
$K_{\bf q}({\bf r,r'}) = K({\bf r,r'}) e^{-i{\bf q}\cdot({\bf r-r'})}$ via 
\begin{equation}
\label{vofrho}
 V_{\bf q \omega}^{\lambda}({\bf r})=\int d^3 r' K_{\bf q}({\bf r,r}') n^{\lambda}_{{\bf q}\omega}({\bf r}').
\end{equation}
Note that $n^\lambda$ and $V^\lambda$ are implicitly understood as four-vectors (or, equivalently, as $2\times 2$ matrices~\cite{RicciPRB19}), as they contain information about both the charge and the magnetization density; for the same reason, the interaction kernel $K$ is generally a $4\times 4$ matrix function of ${\bf r}$ and ${\bf r'}$.
Straightforward application of time-dependent perturbation theory then leads to the first-order variation of the observable $\lambda_a$ in response to the external field $\lambda_b$, 
\begin{equation}
\label{e_ab}
 E_{ab}(\w,{\bf q}) = \int [d^3 k] {\rm Tr} \, [H^a_{\bf k+q,-q} \, \hat{\rho}^b_{\bf k,q}(\w) + H^{ab}_{\bf k}({\bf q}) \hat{\rho}_{\bf k} ].
\end{equation}
Here, we have introduced the second-order Hamiltonian $H^{ab}_{\bf k}({\bf q})$; and the ground-state, $\hat{\rho}_{\bf k}$, and first-order, $\hat{\rho}^b_{\bf k,q}(\w)$, density operators.
The latter describes the response of the electronic system to the modulated external field and also defines the first-order electron density that enters Eq.~(\ref{vofrho}),
\begin{equation}
\label{nb}
	n^\la_{\bf q \w}({\bf r}) = \rm{Tr} \left[ \bm{\sigma} \int [d^3 k] \,
	\langle {\bf r}|\hat{\rho}^\la_{\bf k,q}(\w)|{\bf r}\rangle \right],
\end{equation}
where the trace is taken in spinor space.
Here, $n^{\lambda}=(n^{\lambda}_0,{\bf m}^\lambda)$, with $n^{\lambda}_0$ being the 
first-order particle density and $m_\alpha^\lambda$ the $\alpha=x,\, y,\, z$ component of the 
first-order magnetization density, and $\bm{\sigma}=(\sigma_0,\sigma_x,\sigma_y,\sigma_x)$ is here a four-vector with $\sigma_0$ being the $2 \times 2$ identity matrix and $\sigma_\alpha$ the Pauli matrices.
As the perturbing potential contains a contribution of the response, the linear problem needs to be solved self-consistently.

Following the standard DFPT rules, the first-order density operator can be written as
\begin{equation}
\label{rhob}
	\hat{\rho}^\la_{\bf k,q}(\w) = \sum_n f_{n\bf k} \left( 
	|u^\la_{n\bf k,q}(\w) \rangle \langle u_{n\bf k}| +
	|u_{n\bf k}\rangle \langle u^\la_{n\bf k+q,-q}(-\w)| \right),
\end{equation}
where the sum is carried out on the valence states only ($f_{n\bf k}=1$ are the band occupancies), and first-order spinor wavefunctions are given by the following Sternheimer equation,
\begin{widetext}
\begin{equation}
 \left(\hat{H}_{\bf k+q} + a\hat{P}_{\bf k+q} - \epsilon_{n\bf k} - \omega \right)
|u_{m\bf k,q}^{\lambda}(\omega) \rangle = 
-\hat{Q}_{\bf k+q} \mathcal{H}^{\lambda}_{\bf k,q} (\omega)
|u_{m\bf k}^{(0)} \rangle.
\label{eq:stern_res}
\end{equation}
\end{widetext}
Note that the calculation of the resonant wavefunctions at ${\bf +q,+\w}$ needs to be carried out in parallel with the antiresonant ones at ${\bf -q,-\w}$, as they are mutually coupled via the SCF potential. [The latter is given by the combination of Eq.(\ref{rhob}), (\ref{nb}) and (\ref{vofrho}).] 
\blue{This formulation is in all respects equivalent to existing approaches that are based on the use of the time-reversal operator.~\cite{GorniEPJB18,UrruPRB19}}

By plugging Eq.~\eqref{rhob} into Eq.~(\ref{e_ab}), we can also write the second derivative more explicitly as
\begin{equation}
E_{ab}({\bf q},\omega) =  \bar{E}_{ab}({\bf q},\omega) +
  \bar{E}^*_{ab}(-{\bf q},-\omega) + {E}^{\rm sta}_{ab}({\bf q}),
\end{equation}
where the static (sta) contribution corresponds to the second term on the rhs of Eq.~(\ref{e_ab}), and we have decomposed the dynamical contribution of the first term into resonant and antiresonant parts by defining
\begin{equation}
\label{eq:nonvar_kubo}
\bar{E}_{ab}({\bf q},\omega) = \int [d^3 k] \sum_m
\langle u_{m\bf k} | (H^{a}_{\bf k,q})^\dagger
   |u_{m\bf k,q}^{b}(\omega) \rangle.
\end{equation}
This result generalizes the nonstationary expression for the linear response in the
adiabatic regime~\cite{gonze,gonze/lee}, with the latter being recovered when $\omega=0$.
Eq.~\eqref{eq:nonvar_kubo} allows us to obtain the desired second-order
energies once the first-order response functions to one of the
perturbations ($\lambda_b$) have been calculated, e.g, by iteratively solving
the Sternheimer equations.

\subsection{Treatment of the magnetic variables \label{sec:penalty}}

To differentiate the magnetic functionals of Section~\ref{sec:gsfun} with respect to a perturbation parameter $\lambda$, the central quantity we need to consider are the first-order magnetic moments, ${\bf m}_{\kappa}^\lambda$.
Whenever the functions $f_\kappa$ do not depend explicitly on $\lambda$, 
${\bf m}_{\kappa}^\lambda$ are given by the the integral over the atomic spheres of the first-order magnetization density,
\begin{equation}
\begin{split}
{\bf m}_{\kappa}^\lambda({\bf q},\omega) = &
 \int_{\rm cell} d^3 r \, 
 f_{\kappa}^{\bf -q}({\bf r})  \,
 {\bf m}^\lambda_{\bf q\w}({\bf r}). \end{split}
\label{eq:m1mom}
\end{equation}
Here $\kappa$ runs over the active magnetic sites  
and we have introduced the cell-periodic form factor of the atomic spheres via 
\begin{equation}
f_{\kappa}^{\bf q}({\bf r}) = f_\kappa({\bf r}-\bm{\tau}_\kappa)
e^{i {\bf q}(\bm{\tau}_\kappa-{\bf r})},
\end{equation}
This function enters directly the ``external field'' perturbation associated with a local Zeeman potential, which reads as
\begin{equation}
V^{H_l}_{\bf q}({\bf r}) = - \sigma_\beta
f_{\kappa}^{\bf q}({\bf r}),
\label{eq:v1zeem}
\end{equation}
By plugging this perturbation into the DFPT formalism of the previous Section, we can readily calculate the response of a noncollinear system to a local Zeeman field; this is essentially all we need in the context of the enthalpy functionals ${F}$ and $\tilde{F}$.

Note that, in the specific case where $\lambda=\tau_{\kappa' \beta}$ is a phonon perturbation, there is generally an additional contribution to Eq.~\eqref{eq:m1mom},
\begin{equation}
\begin{split}
\Delta {\bf m}_{\kappa}^{\tau_{\kappa'\beta}}({\bf q},\omega) = &
 - \delta_{\kappa \kappa'} \int_{\rm cell} d^3 r \, 
 \frac{\partial f_{\kappa}({\bf r})}{\partial r_\beta}  \,
 {\bf m}({\bf r}) 
\end{split}
\end{equation}
This second piece is a momentum-independent static contribution, and describes the ground-state magnetization that crosses the boundary of the sphere when the latter is displaced along the phonon trajectory.
This contribution vanishes identically in our CrI$_3$ \blue{and Cr$_2$O$_3$} testcases, and we won't consider it further.

In the penalty-based internal energy $\tilde{U}$, 
the external potential (derivative of the Hamiltonian with respect to the parameter $B_l$) is related to the local Zeeman field via
\begin{equation}
V^{B_l}_{\bf q}({\bf r}) = \alpha V^{H_l}_{\bf q}({\bf r}),
\end{equation}
i.e., the two perturbations are essentially identical except for a trivial factor of $\alpha$.
At difference with the case of the magnetic enthalpy, however, the penalty term in Eq.~\eqref{Utilde} entails a modification of the interaction kernel,
\begin{equation}
	\bar{K}_{\bf q}({\bf r,r'}) = {K}_{\bf q}({\bf r,r'}) + \alpha  \sum_l V^{H_l}_{\bf q}({\bf r})  V^{H_l}_{\bf -q}({\bf r}').
\end{equation}
This means that the total Zeeman field acting on a given sphere, inclusive of the self-consistent contribution of the spin response mediated by the additional term in $\bar{K}$, is given by
\begin{equation}
H^{\lambda}_l = \alpha	 [ \delta_{l \lambda} -m^{\lambda}_l({\bf q},\omega)],
\end{equation}
consistent with Eq.~\eqref{eq:hl}. Note that, similarly to the case of the enthalpy functional $\tilde{F}$, there is an additional nonvariational contribution to the second-order energy. 
It arises in the second derivatives of $\tilde{U}$ with respect to two $B_l$'s, and is in the form $\Delta \tilde{U}_{jl} = \alpha \delta_{jl}$.

To minimize the dependence of the calculated parameters on the atomic sphere functions $f_\kappa({\bf r})$, we follow the prescriptions of Ref.~\onlinecite{RenPRX24} and normalize the definition of all magnetic variables to the amplitude of the ground-state magnetic moments, $m^0_\kappa$. 
Then, the first-order local magnetic moments ${\bf m}^\lambda_\kappa$ and their conjugate Zeeman fields $H^\lambda_l$ acquire the physical meaning of induced canting angles and torques, respectively. 
In practice, we operate the normalization a posteriori on the second-order coefficients, multiplying (dividing) the result by $m^0_\kappa$ every time that the energy is derived with respect to $M_l$ ($H_l$).

\subsection{Stationary formulation and Berry curvatures}

The dynamical linear-response approach described in the previous paragraphs enjoys a 
variational formulation, where the Sternheimer equation of Eq.~(\ref{eq:stern_res})
emerges from a stationary principle imposed on the second-order action functional~\cite{DalCorsoPRB96}.
More specifically, we rewrite the second derivative of the action in the following form,
\begin{equation}
\label{eq:chistat}
\begin{split}
E_{ab}({\bf q},\omega) =  &
 \tilde{E}_{ab}({\bf q},\omega) +
  \tilde{E}^*_{ab}(-{\bf q},-\omega) \\
  &+ E^{\rm SCF}_{ab}({\bf q},\omega) + {E}^{\rm sta}_{ab}({\bf q}),
\end{split}
\end{equation}
where the tilded symbols correspond to
\begin{equation}
\label{eq:tildechi}
\begin{split}
&  \tilde{E}_{ab}({\bf q},\omega) =   \int[d^3 k] \sum_m \Big[ \\
&\qquad \langle {u}_{m\bf k,q}^{a}(\omega)|
 \left(H_{\bf k+q}  - \epsilon_{n\bf k} - \omega \right)
|u_{m\bf k,q}^{b}(\omega) \rangle \\
& \qquad + \langle {u}_{m\bf k,q}^{a}(\omega)|  H^{b}_{\bf k,q}
|u_{m\bf k}^{(0)} \rangle \\
& \qquad + \langle u_{m\bf k}^{(0)} | (H^{a}_{\bf k,q})^\dagger  | u_{m\bf k,q}^{b}(\omega)\rangle \Big],
\end{split}
\end{equation}
and the new SCF contribution is
\begin{equation}
E^{\rm SCF}_{ab}({\bf q},\omega) = \int d^3 r \int d^3 r' [{n}^{a}_{\bf q\omega}]^* ({\bf r})
 K_{\bf q}({\bf r},{\bf r'}) n^{b}_{\bf q\omega} ({\bf r}').
\end{equation}
$E_{ab}({\bf q},\omega)$ is intended to be minimized while imposing the parallel-transport gauge on the first-order wave functions,
which requires
\begin{equation}
\label{partrans}
  \langle u_{m\bf k+q}    | u_{n\bf k,q}^{\lambda}(\omega)\rangle = 0
\end{equation}
for all $m$ and $n$ running over the valence manifold. [Alternatively, one can avoid enforcing Eq.~\eqref{partrans} explicitly by adopting 
the unconstrained approach of Ref.~\onlinecite{Royo2019}.] It is straightforward to verify that the stationary condition on $E_{ab}({\bf q},\omega)$
yields the Sternheimer equation~\eqref{eq:stern_res}.

The stationary nature of the second-order functional, represented in Eq.~\eqref{eq:chistat}, 
greatly facilitates a direct computation of higher-order parametric derivatives 
thanks to the $2n+1$ theorem.
Here, we specifically examine the case of the frequency derivative, which reads as 
\begin{equation}
\begin{split}
& \frac{d E_{ab}({\bf q},\omega)}{d\omega}=
\frac{\partial E_{ab}({\bf q},\omega)}{\partial\omega}= \int [d^3 k] \sum_m f_{n\bf k} \\
& \quad \left( \langle u_{m\bf k,-q}^b(-\omega) | {u}_{m\bf k,-q}^a(-\omega) \rangle - \langle {u}_{m\bf k,q}^a(\omega)|u_{m\bf k,q}^b(\omega) \rangle
 \right).
\end{split}
\label{eq:berry}
\end{equation}
In the static limit and at the zone center, this formula takes the 
form of a Berry curvature calculated in the parameter space spanned by $\lambda_a$ and $\lambda_b$;
Eq.~(\ref{eq:berry}) can be regarded as the finite-momentum and finite-frequency generalization thereof.
In the special case of two magnetic perturbations, our Eq.~(\ref{eq:berry})
recovers Eq.(14b) of Lin and Feng~\cite{lin2025}.
Our theory, however, is applicable to any pair of perturbations of
arbitrary physical nature; it provides therefore a natural methodological 
framework for, e.g., the approach to coupled spin-lattice dynamics proposed
in Ref.~\onlinecite{RenPRX24}.

\begin{figure}
 \centering
 \includegraphics[width=0.8\columnwidth]{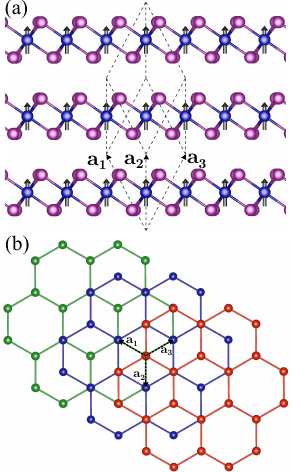}
 \caption{(a) Side view and (b) top view of a three-layer section of the CrI$_3$ structure. The dashed lines in (a) outline the primitive cell used in the numerical calculations, with the three real-space primitive vectors labeled as ${\bf a}_i$. Thick black arrows indicate the magnetic moments, aligned along the $z$-axis. In panel (b), only the Cr sites are shown to clearly illustrate the ABC hexagonal stacking pattern.}
 \label{fig:structure}
\end{figure}

\section{Computational details and validation \label{sec:validate}}

\begin{figure}
 \centering
 \includegraphics[width=0.4\columnwidth]{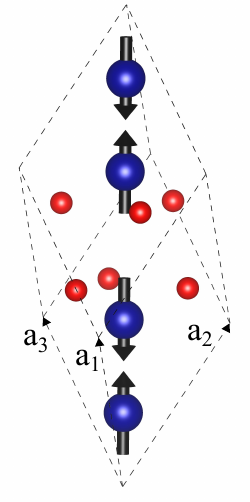}
 \caption{Side view of the Cr$_2$O$_3$ rhombohedral cell. The dashed lines outline the three real-space primitive vectors, labeled as ${\bf a}_i$. Thick black arrows indicate the magnetic moments, aligned along the $z$-axis.}
 \label{fig:Cr2O3_structure}
\end{figure}

\subsection{Computational details \label{sec:cparam}}

\begin{figure*}
 \centering
 \includegraphics[width=1.0\textwidth]{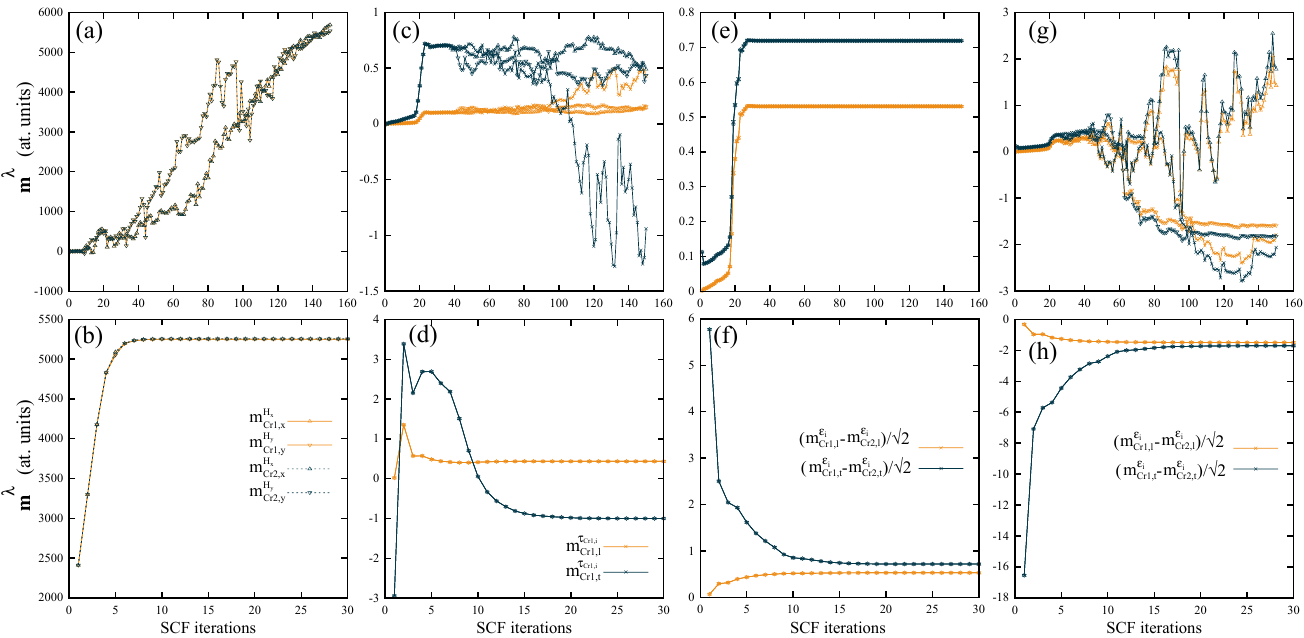}
 \caption{Convergence rate of first-order local magnetic moments versus number of SCF iterations
 within standard noncollinear DFPT (top panels) and within our constrained-${\bf B}$ method (bottom panels).
The following types of perturbations are considered: static uniform Zeeman field (a,b),
 static displacement of Cr1 (c,d), static electric field (e,f)
 and dynamic electric field at $\hbar\omega=32.7$ meV (g,h). In the latter case, only
 the real part of the magnetic moment is shown. 
 The first-order magnetic moments are 
 expressed in Cartesian coordinates in panels (a,b). 
In (c,d), we show their projection on a direction that is either parallel (l) or perpendicular (t) to the atomic displacement in the $xy$ plane.  In (e,f,g,h) the antiferromagnetic component of the induced moments is projected on the applied electric field direction (l) or its normal (t).}
 \label{fig:convergence}
\end{figure*}

We perform our calculations using noncollinear DFT~\cite{Gonze2016} and DFPT~\cite{RicciPRB19} as implemented in ABINIT v9.9,
where we have incorporated the dynamical approach at constrained magnetic moments described in Sections~\ref{sec:theory} and~\ref{sec:implement}.
The exchange-correlation kernel is described using the Perdew and Wang parametrization~\cite{perdew/wang:1992} of the Local Density Approximation (LDA) for both the ground state and linear-response calculations.
We use full relativistic norm-conserving pseudopotentials from Pseudo Dojo~\cite{pseudodojo},
regenerating them for LDA using the ONCVPSP software~\cite{Hamann1979,Hamann1989}.
The calculations employ a plane-wave cutoff of 40 Ha, \blue{a hexagonal $6\times6\times6$} {\bf k}-point mesh, and incorporate spin-orbit coupling throughout\blue{~\cite{Gonze2002,VerstraetePRB08}}.
\blue{
We note that, at the time of writing, the treatment of crystal symmetries in magnetic systems is still unreliable in ABINIT;
therefore, all our calculations are performed without taking advantage of magnetic space groups to reduce, e.g., the
number of $k$-points and independent phonon/spin perturbations.}

The dimensions of the rhombohedral CrI$_3$ primitive cell are fixed to the experimental values~\cite{McGuireCM15},
with atomic coordinates optimized until all forces are smaller than $10^{-6}$ Ha/bohr.
Unless otherwise stated, we set the atomic sphere radius to $R^{\rm WS}=2$ bohr for all atoms, and the boundary smearing width to 0.5 bohr.
This set up correctly predicts a ferromagnetic ground state with a total magnetic moment of $|m^0|\sim6.02$ $\mu_{\rm B}$
oriented along the out-of-plane direction $z$. 
In each CrI$_3$ layer, the Cr atoms form a honeycomb lattice
with two sites that are related by a space
inversion symmetry operation (see Fig~\ref{fig:structure} (b)).
The structure is slightly buckled, with the individual
Cr's lying either above or below the central plane of
each layer; they will be referred to respectively as
``Cr1'' and ``Cr2'' henceforth.
The calculated local magnetic moments in the ground state
amount to $|m^0_{\rm Cr}|\sim2.54$ $\mu_{\rm B}$.

\blue{For Cr$_2$O$_3$, standard DFT calculations predict a ground state with an easy-plane magnetic anisotropy, in contrast to the experimentally observed easy axis along the out-of-plane, [111] direction of the rhombohedral primitive cell~\cite{MuPRM19}.
To correct for this discrepancy, we applied a 2\% epitaxial strain in the in-plane directions while allowing relaxation along the out-of-plane axis. This yields a rhombohedral lattice with $a = 5.358$ \AA~and $\alpha = 58.65^\circ$.
In this configuration, the system reproduces the experimental easy-axis alignment shown in Fig.~\ref{fig:Cr2O3_structure}, where the Cr magnetic moments are chosen to form an in-pointing antiferromagnetic arrangement along the $z$ axis.
The Cr ions will be referred as ``Cr1, Cr2, Cr3 and Cr4'' according to their position along the axis z, i.e., from bottom to top their ground-state magnetic moments point up, down, up and down.
The corresponding local magnetic moments, obtained within the aforementioned atomic spheres, are $|m_{\rm Cr}^0|=2.23$ $\mu_{\rm B}$.
}

In the linear-response calculations, we use a a  penalty parameter of $\alpha=0.0125$ Ha, with a
stringent tolerance criterion of $10^{-14}$ for the squared residual
of the first-order potential.
To obtain linear-response quantities over a broad and dense frequency mesh, we first perform
constrained-${\bf B}$ DFPT calculations over a coarse grid of $\hbar\omega=$0.0, 0.6, 1.2,
1.8 and 2.4 mHa. 
(Given that all coefficients are smooth functions of $\omega$, an even coarser mesh could
be used without losing accuracy.)
At each frequency, we then convert the result into the frozen-spin (FS) flavor of the constrained-spin response.
We then we perform a fourth-order polynomial interpolation thereof and operate the analytic continuation of Eq.~\eqref{u_causal} with a value of $\eta=1$ $\mu$Ha.
Finally, we calculate the correlation and susceptibility functions of Secs.~\ref{sec:corrfun} and~\ref{sec:sus}, respectively, over the vibrational and magnonic frequency ranges.

\subsection{Performance of the algorithm}

In this section, we demonstrate the practical advantages of our implementation of DFPT at constrained magnetic moments,
as opposed to the standard approach to the linear response of a noncollinear magnet.
To that end, \blue{by using CrI$_3$ as a testcase}, we study the evolution of the first-order magnetic moments ($m_{\kappa\alpha}^\lambda$) throughout the self-consistent iterations,
and use them as indicators of the convergence rate of the calculation in both the static and dynamic regimes.
We illustrate the difficulties in converging $m_{\kappa\alpha}^\lambda$ within standard DFPT first.

Figs.~\ref{fig:convergence}(a,c) show the 
evolution of the $m_{\kappa\alpha}^\lambda$
induced by a uniform Zeeman field and the displacements of Cr1, respectively.
Not surprisingly, the uniform in-plane Zeeman field produces the strongest magnetic response,
consisting in an equally large canting of both ${\bf m}_{\rm Cr1}$ and ${\bf m}_{\rm Cr2}$ along the field direction.
As shown in Fig.~\ref{fig:convergence}(a), however, the values remain far from converged even
after numerous self-consistent iterations.
These numerical difficulties are due to the acoustic magnon resonance, which
occurs at 0.74 meV within our computational model and couples strongly with the Zeeman field.
Such an exceedingly small energy is a consequence of the weak magnetocrystalline anisotropy, which
depends on spin-orbit coupling.
This low-lying electronic excitation, in turn, makes the linear-response problem ill-conditioned,
making convergence nearly impossible.
A similar catastrophic behavior occurs in the case of the phonon perturbation [Fig.~\ref{fig:convergence}(c)],
where not only is convergence unattained, but the expected symmetry in the response along the three reduced directions
breaks down during the self-consistent process, suggesting a nearly chaotic regime.

\begin{table}
\setlength{\tabcolsep}{3pt}
\begin{center}
\caption{\label{tab:invariance} Selected coefficients of the CI local spin susceptibility and the CI and RS dielectric
tensor ($\varepsilon_{xy}$) calculated at $\hbar\omega=32.7$ meV and $\eta=0$ with different values of the penalty parameter ($\alpha$)
and radius of the WS atomic spheres ($R^{\rm WS}$). All quantities are expressed in atomic units and only the relevant real or imaginary part of the complex coefficients is shown.}

\begin{tabular}{cc|rrrr}\hline\hline
   \T\B $\alpha\times10^3$& $R^{\rm WS}$ & \multicolumn{1}{c}{$\chi_{11}'$} & \multicolumn{1}{c}{$\chi_{13}''$} &
   \multicolumn{1}{c}{$\varepsilon_{xx}'$} &  \multicolumn{1}{c}{$\varepsilon_{xy}''$}  \\ 
   \hline
   \T\B 6.25 & 2.0 & $-$57.8669 & $-$83.9676  &
                                             8.80455   & $-$0.00162 \\
   \T\B 12.50 & 2.0 & $-$57.8669 & $-$83.9676 &
                                             8.80455   & $-$0.00162 \\
   \T\B 17.50 & 2.0 & $-$57.8668 & $-$83.9675 &
                                             8.80455   & $-$0.00162 \\
   \T\B 12.50 & 2.5 & $-$57.4413 & $-$83.4718 &
                                             8.80455   & $-$0.00162 \\
  \hline \hline
\end{tabular}
\end{center}
\end{table}

Figs.~\ref{fig:convergence}(e,g) 
show the convergence of the magnetic moments induced by a static
and dynamic ($\hbar\omega=32.7$ meV) electric field, respectively.
Interestingly, in the static case convergence is achieved after a reasonable number of iterations, at difference with the previous examples.
This is because an electric field cannot couple by symmetry
to the acoustic magnon in CrI$_3$.
Instead, it couples to the optical magnon (involving opposite spin canting of the two Cr ions), which occurs at a much higher frequency (28.27 meV) in our calculations, and hence does not affect the condition number of the problem as dramatically.
Notwithstanding, the conditioning issues are back in force if we probe the system dynamically, for instance at a frequency that is slightly larger than the magnon resonance [Fig.~\ref{fig:structure}(g)].

All the above difficulties are resolved in full by our constrained-${\bf B}$ approach.
In the bottom panels of Fig.~\ref{fig:structure} we show the results for the first-order magnetic moments
after converting them from their fixed-spin to the physical relaxed-spin values. 
In all cases, full convergence is achieved after about 20 iterations, with no hint of 
the erratic behavior observed in the above panels, and
a tolerance that can be
systematically improved to essentially machine precision.
Additionally, a comparison of the results for the static electric field obtained via both approaches [Figs.~\ref{fig:convergence} (e) and (f)] reveals that, upon convergence, the values differ only at the fifth decimal place. This agreement validates the robustness of our formalism and computational implementation.
Overall, these results demonstrate that our method brings a
noncollinear spin calculation to the same level of computational efficiency as that of a standard
nonmagnetic insulator.

\subsection{Influence of the penalty parameters}

Here, we asses the influence of the parameters defining the magnetic penalty --the stiffness
($\alpha$), and the radius ($R^{\rm WS}$) of the WS atomic spheres-- in terms of computational
efficiency and impact on the final results.
Table~\ref{tab:invariance} presents a set of selected coefficients of the CI local spin
susceptibility and the CI and RS dielectric tensor calculated with different combinations of $\alpha$ and $R^{\rm WS}$,
while keeping the smearing constant.
The constrained-${\bf B}$ values of these second-order tensors quantitatively depend on the penalty
parameters (not shown). 
However, the macroscopic observables at relaxed spins (constrained-${\bf H}$) are essentially unaffected by this choice, as
illustrated in the table for the dielectric tensor.

The local spin susceptibility is not macroscopic, meaning it is not averaged over the entire unit cell but rather over the WS spheres.
This is why the magnitude of its coefficients shows a slight dependence on $R^{\rm WS}$ (but not on the penalty parameter,
since ${\chi}_{jl}$ is the inverse of the fixed-spin $U_{jl}$ matrix).
Importantly, the extracted magnon frequencies are independent on both
parameters, as they should.
[Note that the negative sign of $\chi_{11}'$ in Tab.~\ref{tab:invariance} is physically correct, as it is evaluated
at a frequency higher than the optical magnon resonance (see, e.g., Fig.~\ref{fig:magsus}).]

\begin{figure}
 \centering
 \includegraphics[width=1.0\columnwidth]{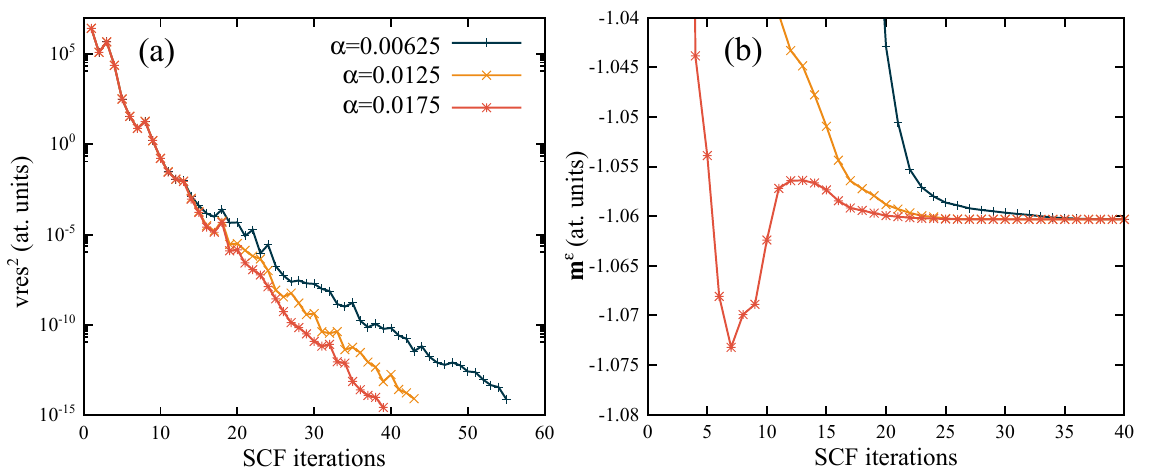}
 \caption{(a) Squared residual of the SCF potential and (b) longitudinal in-plane first-order atomic magnetic moments as a function of the SCF iterations for the linear-response calculation to an electric field of $\hbar\omega=32.7$ meV. Colored curves illustrate the results obtained with different values of the magnetic penalty stiffness ($\alpha$).}
 \label{fig:vres}
\end{figure}

Since the final results are independent on the penalty parameter $\alpha$, the latter can be tuned to maximize computational efficiency. For instance, larger values of $\alpha$ result in stiffer spin modes, which facilitate smoother and faster convergence to a fixed tolerance by reducing the condition number of the problem.
This is illustrated in Fig.~\ref{fig:vres} for the linear-response to a finite-frequency electric field. 
The values of $\alpha = 0.00625$,  $0.0125$ and $0.0175$ Ha used in our calculations respectively shift the magnon resonances to energies of approximately 0.7, 1.4 and 2.0 eV, namely, of the order of and well above the electronic band gap ($\sim 0.9$ eV as per our LDA calculation for the direct transition).
Using even larger values of $\alpha$ 
would be ineffective, as the condition number is bounded by the onset of cross-gap electronic excitations.
This fact can be appreciated in Fig.~\ref{fig:vres} (b) where the first-order magnetic moments obtained for $\alpha=0.0125$ and $0.0175$ Ha reach the converged value in approximately the same number of iterations.

Having demonstrated the performance of the constrained-${\bf B}$ approach, in what follows, we shall use it to study the physical linear-response properties of bulk CrI3 \blue{and Cr$_2$O$_3$} associated with external electric and Zeeman fields, as well as lattice displacements.

\begin{table}
\setlength{\tabcolsep}{4pt}
\begin{center}
\caption{\label{tab:exp_param} Parameters of the adiabatic expansion [Eq.~\eqref{eq:linearw}] for the matrix $U^{(ss)}$ expressed in the basis of acoustic and optical magnons. $K$ and $G$ are analytically calculated from Eqs.~\eqref{ujl} and~\eqref{eq:berry}, respectively, and $M$ is numerically extracted by fitting the $n=4$ data of Fig.~\ref{fig:lineint} (the same fitting provides $G$ numbers that agree with the analytical ones up to the fifth decimal position). Values between brackets are extracted from Ref.~\onlinecite{RenPRX24}.}

\begin{tabular}{c|rrc}\hline\hline
   \T\B & \multicolumn{1}{c}{$K^{(\rm ss)}$ (meV)} & \multicolumn{1}{c}{$G^{(\rm ss)}$} &
   \multicolumn{1}{c}{$M^{(\rm ss)}$ (eV$^{-1}$)}  \\
   \hline
   \T\B AM &   1.112 (0.885)  &  1.500 (1.500) & 0.034 \\
   \T\B OM &  43.865 (35.795) &  1.545 (1.566) & 0.244 \\
   \hline \hline
\end{tabular}
\end{center}
\end{table}

\section{Ferromagnetic C\lowercase{r}I$_3$ \label{sec:CrI3}}

\blue{CrI$_3$ is a van der Waals ferromagnet that remains ordered down to the monolayer limit~\cite{FuangNAT17}.
Its properties arise from the interplay of electronic, vibrational, and magnetic degrees of freedom~\cite{HuangNN18,SunNAT19,ZhangNM22,EdstromPRL22}, and its spin-wave dispersion has attracted particular interest due to possible nontrivial topology, including chiral magnon edge states~\cite{ChenPRX18,CenkerNat20,BrehmPRB24}.
First-principles studies of magnons in CrI$_3$ have therefore focused mainly on topological aspects~\cite{Gorni_bulk_PRB23}, while magnon-phonon coupling has often been treated with model Hamiltonians~\cite{delugas_mono_PRB23}.
A fully ab initio description was first reported in Ref.~\cite{RenPRX24} using an adiabatic approximation equivalent to the FOA discussed below, which revealed magnon-phonon-induced splittings of a few tens of $\mu$eV between circularly polarized chiral phonons.}

\blue{In this section, we benchmark the validity of the adiabatic formalism by comparing it with exact TD-DFPT calculations of bare magnons and hybrid magnon-phonon excitations, and then examine how their dynamics shape the electromagnetic susceptibilities of CrI$_3$.}

\subsection{Bare magnons and adiabatic approximation \label{sec:locmagsus}}

To start with, we study the local spin susceptibility at the CI level as defined in Eqs.~\eqref{chi1} and~\eqref{chi2}.
\blue{The magnetic degrees of freedom span a four-dimensional space (Cr1$_x$,Cr1$_y$,Cr2$_x$,Cr2$_y$),
corresponding to the in-plane components of the Cr magnetic moments or local Zeeman fields.}
To simplify the physical analysis in terms of magnon-like 
normal modes, we first convert the Cartesian $4\times4$ tensor into a pair of $2\times2$ tensors by projecting $\chi_{ij}$
onto the basis of the spin cantings (or Zeeman fields) associated with the acoustic and the optical magnon,~\cite{RenPRX24}
\begin{equation}
 {\bf e^{\rm AM}}=\frac{1}{\sqrt{2}}
 \left( \begin{matrix}
 1 & 0 \\
 0 & 1 \\
 1 & 0 \\
 0 & 1
 \end{matrix} \right) , \quad 
 {\bf e^{\rm OM}}=\frac{1}{\sqrt{2}}
 \left( \begin{matrix}
 1 & 0 \\
 0 & 1 \\
 -1 & 0 \\
 0 & -1
 \end{matrix} \right).
 \label{eq:magnon_basis}
\end{equation}
In such a basis the inverse susceptibility ${\bf U}^{(ss)}$ is block-diagonal (the AM and OM sectors are decoupled),
and takes the following form,
\begin{equation}
\label{bfu}
{\bf U}^{(ss)} = \begin{pmatrix}
{\bf u}^{\rm AM} & 0 \\
0 & {\bf u}^{\rm OM}
\end{pmatrix}, \qquad
{\bf u} = \begin{pmatrix}
a & ib \\
-ib & a
\end{pmatrix}
\end{equation}
Here $a=K - \w^2 M + \cdots$ and $b = \w G + \cdots$ are scalar real functions of $\w$
 such that $a(-\w)=a(\w)$ and $b(-\w)=-b(\w)$.

Next, we project the result onto the basis of circularly polarized modes, ${\bf \hat{e}}_\pm=1/\sqrt{2}({\bf \hat{e}}_x \mp i {\bf \hat{e}}_y)$, where ${\bf \hat{e}}_{x/y}$ are the Cartesian basis vectors. 
Both ${\bf u}^{\rm AM}$ and ${\bf u}^{\rm OM}$ become then diagonal, with eigenvalues
\begin{equation}
\label{u_pm}
u_{\pm} = a(\w) \mp b(\w) \simeq K \mp \w G - \w^2 M.
\end{equation}
The zeros of $u_\pm$ correspond to the magnon resonances.
The ``$+$'' and ``$-$'' modes correspond to clockwise ($+$) and counterclockwise ($-$) spin precession, respectively, within the $xy$ plane [Fig.~\ref{fig:structure}].
The calculated parameters of the adiabatic expansion of Eq.~\eqref{u_pm} are reported in Table~\ref{tab:exp_param}.
Within this setup, we find the AM and OM resonances at 0.74 meV and 28.27 meV, respectively.
The overestimation of the latter compared to the experimental value --about 19 meV in Ref.~\cite{ChenPRX18}-- is consistent
with previous theoretical studies using similar DFPT approaches,~\cite{Gorni_bulk_PRB23}.
Inclusion of terms in Eq.~\eqref{u_pm} beyond $O(\w^2)$ does not affect the calculated frequencies within machine precision: this result confirms that 
the SOA (see Sec.~\ref{sec:linearw}) is essentially exact in bulk CrI$_3$.

\begin{figure}
 \centering
 \includegraphics[width=1.0\columnwidth]{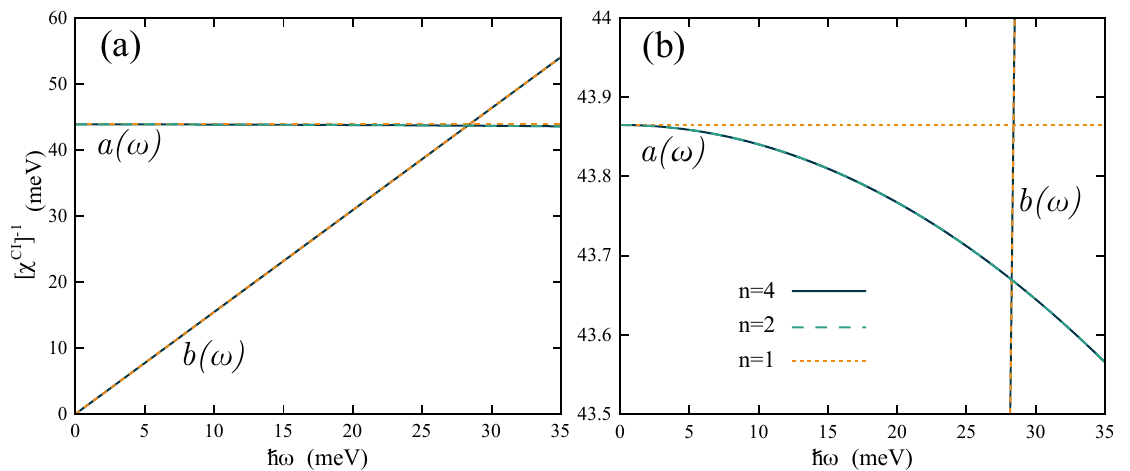}
 \caption{(a) Frequency dependence of the optical-magnon coefficients of Eq.~\eqref{bfu}, as obtained from polynomial frequency interpolation at different degrees: $n=4$ (dark-solid), $n=2$ (turquoise-dashed) and $n=1$ (orange-dashed).
 Panel (b) shows an amplified view of the abscissas axis around the intersection of the curves shown in (a).}
 \label{fig:lineint}
\end{figure}

We are now in a good position to test the performance of the FOA level of theory [Sec.~\ref{sec:linearw}] that has oftentimes been adopted in the literature~\cite{BoniniPRL23,RenPRX24,lin2025}.
The magnon frequencies are given by $K/G$ within FOA: by using the values of Table~\ref{tab:exp_param} we obtain 0.74 meV and 28.39 meV for the AM and OM, respectively.
In the case of the acoustic magnon, the FOA values match the exact ones within a remarkable  
accuracy of one part in $10^{-5}$ meV, well below other sources of numerical error. 
Conversely, the bare optical magnon displays a significant discrepancy, 
with a resonance that appears 0.12 meV blue-shifted at the FOA level.
To explain the reason,
in Figs.~\ref{fig:lineint} (a-b) we plot the $a^{\rm OM}(\w)$ and $b^{\rm OM}(\w)$ functions of Eq.~\eqref{bfu}.
The FOA appears as an excellent approximation for $b(\w)$, which exhibits an almost exactly linear dispersion within the relevant frequency range. On the other hand, $a(\w)$ shows a parabolic dispersion that is missed by the FOA, since the latter assumes a constant $a^{\rm FOA}(\w)=K$.
Within the SOA, the inclusion of an effective mass coefficient via $a(\w)=K-M\w^2$ corrects this limitation, yielding an essentially exact description of $a(\w)$. 

Summarizing this section, our second-order adiabatic approximation constitutes a robust improvement over earlier semiclassical treatments of spin dynamics. It also avoids the necessity for the iterative refinement of the Hessian and Berry curvature coefficients that was proposed recently by Lin and Feng~\cite{lin2025}.

\subsection{Coupled spin-phonon dynamics \label{sec:lattdyn}}

We now assess the impact of the dynamical spin-lattice coupling on the magnon and phonon frequency spectrum of CrI$_3$. 
We focus on the Brillouin zone (BZ) center, and defer the analysis of the finite-${\bf q}$ case to a forthcoming publication.
The nonanalytic dipole-dipole terms are neglected, as they are largely irrelevant to what follows.

To start with, it is useful to recall the multiplet structure that is associated with the space group of the crystal. Space inversion symmetry implies a first partition into gerade ($g$) and ungerade ($u$) modes, which can be further decomposed into sets of one-dimensional (4$A_{u}$ and 4$A_{g}$) and two-dimensional (4$E_{u}$ and 4$E_{g}$) irreps.
Due to the broken time-reversal symmetry (TRS) environment, both $E_{u}$ and $E_{g}$ doublets further split into one-dimensional irreps of opposite chirality ($+$ and $-$), which means that in CrI$_3$ all optical modes 
are singly degenerate~\cite{BoniniPRL23, RenPRX24}.
We find that such splittings are in most cases negligible, consistent with the conclusions of Ref.~\onlinecite{RenPRX24}.
The (highest) $E_u$ doublet at $\sim27$ meV is an interesting exception, in that it interacts strongly with the nearby optical magnon (OM) mode (at 28.43 meV), and provides a representative example of the physics of interest in our context. In the remainder of this Section we shall focus exclusively on this group of three modes,
which we indicate as $E_u^+$, $E_u^-$ and OM henceforth.

\begin{table}
\setlength{\tabcolsep}{7pt}
\begin{center}
\caption{\label{tab:eigenvals} Eigenfrequencies (in meV) of the optical magnon (OM) and the two nearby $E_u^{\pm}$ phonon modes.
Interacting (noninteracting) frequencies are obtained from the poles of the full spin-phonon correlation matrix, ${\bf U}^{-1}$ (the bare magnon, $[{\bf U}^{(ss)}]^{-1}$, and phonon, $[{\bf U}^{(pp)}]^{-1}$, sectors).
Values between brackets are calculated within the FOA of Sec.~\ref{sec:linearw}.
For comparison, the Born-Oppenheimer eigenfrequencies of the phonon doublet calculated at the FS and RS level are: $w^{\rm FS}(E_u)=26.799$ and $w^{\rm RS}(E_u)= 26.789$ meV.}
\begin{tabular}{c|rr}\hline\hline
   \T\B & \multicolumn{1}{c}{Noninteracting} & \multicolumn{1}{c}{Interacting}  \\
   \hline
   \T\B $E_u^+$ &  26.791 (26.794)  &   26.622 (26.635)  \\
   \T\B $E_u^-$ &  26.801 (26.804) &    26.795 (26.800)  \\
   \T\B OM  &  28.271 (28.385) &    28.432 (28.548)  \\
   \hline \hline
\end{tabular}
\end{center}
\end{table}

On the first column of Table~\ref{tab:eigenvals} we show the noninteracting
phonon frequencies, which we calculate by searching for the poles of the bare (FS) propagator, $[{\bf U}^{(pp)}]^{-1}$.
(We also show the noninteracting OM frequency, corresponding to the bare magnon value discussed in the previous Section.)
At this level, the $E_u$ splitting is negligible: $\w(E_u^+)$ and $\w(E_u^-)$ essentially coincide with the Born-Oppenheimer value $\w_0$, indicating that, at frozen spins, the nonadiabaticity of the interatomic force constants is insignificant in this system.
In particular, the FS Berry curvature in phonon space only results in a correction of $\Delta \w(E_u^\pm) = \mp 0.005$ meV, while the mass renormalization due to electron inertia yields an even smaller red-shift of $-0.003$ meV for both $E_u$ modes.

On the rhs of Table~\ref{tab:eigenvals} we show the interacting frequencies, which we define and calculate 
as the poles of the full spin-phonon correlation matrix, ${\bf U}^{-1}$.
Here the $E_u$ doublet presents a sizeable splitting, $\sim -0.17$ meV, between the clockwise ($+$) and counterclockwise ($-$) mode, as $E_u^+$ shifts to lower frequencies compared to $\w_0$.
Such a shift is due to the coupling of $E^+_u$ with the nearby optical magnon, which in turn displays a comparable (0.16 meV) shift to higher frequencies.
Conversely, the $E^-_u$ mode remains largely unaffected, as it couples with the far-away counterclockwise magnon occurring at negative frequencies.
This analysis is in qualitative agreement with that of Ref.~\onlinecite{RenPRX24}, except that in our present computational model the OM lies \emph{above} 
the $E_u$ phonon frequency; hence, the frequency shifts of both $E_u^+$ and the optical magnon are reversed.
As in the noninteracting case, the main difference between the FOA and SOA results appears to reside in the treatment of the bare OM resonance (see the previous Section).

\begin{figure}
 \centering
 \includegraphics[width=1.0\columnwidth]{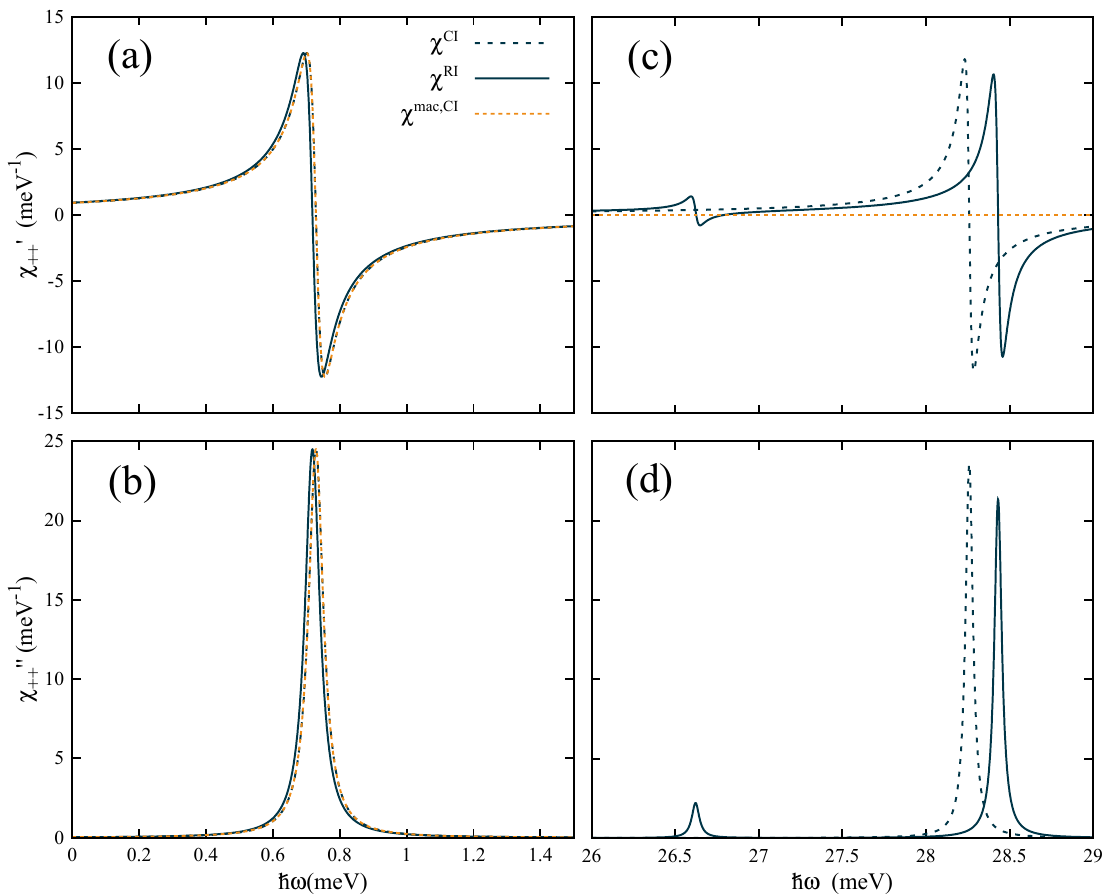}
 \caption{Real (top) and imaginary (bottom) parts of the  local and macroscopic 
 spin susceptibilities as a function of the frequency. Left (right) panels show frequency domains
 around the AM (OM) resonance and dashed (solid) curves are used to represent the CI
 (RI) flavour of the susceptibilities.}
 \label{fig:magsus}
\end{figure}

\subsection{Magnetic susceptibility \label{sec:magsus}}

To illustrate the impact of the $E_u-$OM coupling on the physical properties of the crystal, in Fig.~\ref{fig:magsus} (solid curves) we plot our results for the relaxed-ion susceptibility, ${\bm \chi}^{\rm RI}$.
The main difference with its CI counterpart (dashed curves) consists in a slight shift of the main magnon resonances. The effect is strongest in the case of the optical magnon, where a secondary peak also appears at a frequency of the $E_u^+$-derived mode.
These are both direct consequences of the dynamical spin-phonon interaction discussed above, whereby the magnons hybridize with circularly polarized lattice modes of the same symmetry.
The two peaks at 28.43 and 26.62 meV correspond, respectively, to the predominantly optical magnon-like and $E_u$ phonon-like hybrids.

The impact of lattice relaxation on the acoustic magnon frequency is comparatively much smaller, with a shift of $-0.01$ meV, indicating a weaker coupling to the $E_g$ lattice modes. 
For comparison, in Fig.~\ref{fig:magsus} we also show the CI \emph{macroscopic} spin susceptibility (orange-dashed curves), which is derived from the linear response to a uniform Zeeman field.
(For consistency, we also express the latter as a canting angle in response to a torque, by dividing the calculated susceptibility by a factor of $|m^0|^2$.)
Since, at the considered Brillouin zone center, the uniform field couples only to the acoustic magnon, the susceptibility exhibits a single resonance at 0.73 meV, which perfectly matches the value obtained via the local Zeeman perturbation.
Note that the respective amplitudes of $\chi^{\rm CI}(\w)$ and $\chi^{\rm mac,CI}(\w)$ closely match as well once we take into account a factor of two on the macroscopic susceptibility, which originates from the $1/\sqrt{2}$ normalization factor of the acoustic magnon coordinate, see Eq.~\eqref{eq:magnon_basis}.

\begin{figure}
 \centering
 \includegraphics[width=1.0\columnwidth]{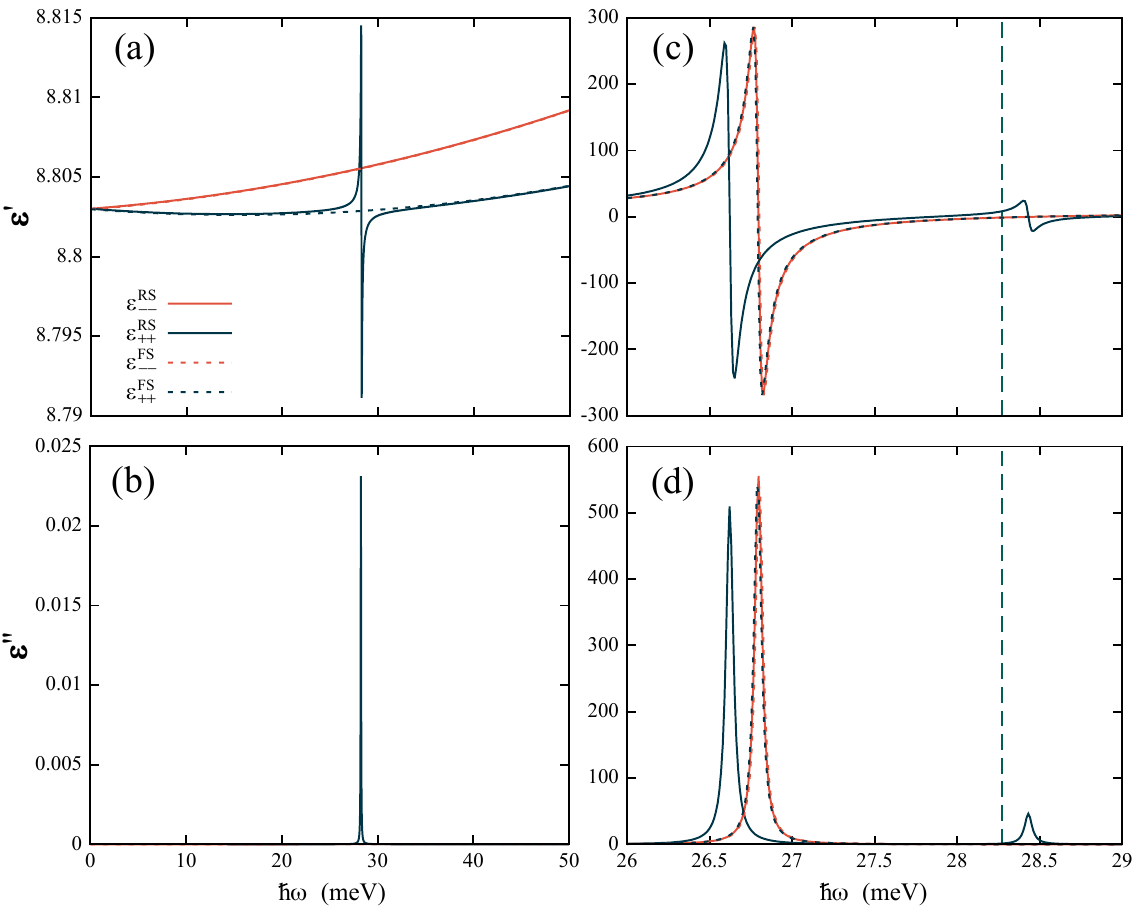}
 \caption{Real (top panels) and imaginary (bottom panels) parts of the diagonal components of the relaxed-spin (RS) and fixed-spin (FS) dielectric tensor in the circularly polarized basis as a function of the frequency. Left (right) panels show the CI (RI) flavour of the tensor. 
 The vertical grey-dashed line depicts the exact position of the CI optical magnon resonance.}
 \label{fig:dielectric}
\end{figure}

\subsection{Dielectric tensor \label{sec:dieltens}}

Next, we investigate the dynamical response of CrI$_3$ to a uniform electric field, which we discuss in terms of the
frequency-dependent dielectric tensor of Eq.~\eqref{eq:epsdie}.
We focus on the in-plane components, which we represent in the circularly polarized $(+,-)$ basis of normal modes.
Our results for $\epsilon_{++}(\w)$ and $\epsilon_{--}(\w)$ are shown in Fig.~\ref{fig:dielectric}. In order to illustrate the respective role played by spin and lattice modes, we repeat our calculations for all four combinations of freezing or relaxing each set of degrees of freedom.

At the FS/CI level [dashed curves in Figs.~\ref{fig:dielectric}(a-b)], both components show a smooth dispersion across a broad frequency
range, and are accurately described by the following quadratic function of $\omega$,
\begin{equation}
\bar{\epsilon}_{\pm\pm}(\w) \simeq \bar{\epsilon} \mp \w \bar{\delta} + \w^2 \bar{\gamma}.
\end{equation}
Our calculated values for these parameters are $\bar{\epsilon} = 8.803$, $\bar{\delta} = 0.047$ eV$^{-1}$ and $\bar{\gamma}=1.532$ eV$^{-2}$.
Upon switching on the spin response while keeping the lattice frozen, an additional contribution appears in the form of a sharp resonance at the bare OM frequency.
This feature can be described (see Sec.~\ref{sec:sus})  as
\begin{equation}
\Delta \epsilon (\w)= \frac{4\pi}{\Omega} \frac{|H^{\mathcal{E}}_+|^2}{K\mp \w G - \w^2 M},
\end{equation}
where the torque on the OM mode in response to an electric field, $H^{\mathcal{E}}_+$, plays
the role of a Born dynamical charge associated with the OM.
The presence of this feature suggests that the optical magnon could, in principle, be detected experimentally via infrared spectroscopy. 
However, the signal is both narrow and weak, potentially making it challenging to resolve as previously noted by Olsen~\cite{OlsenPRL21}.
In the static limit, $\Delta \epsilon (0) = 2.8\times 10^{-5}$, confirming the negligibly small dipolar strength of the bare magnon.

Upon inclusion of the lattice-relaxation contributions, the above picture changes rather drastically.
In the infrared range, the dielectric response is dominated by the strong resonances associated with the $E_u$ phonons. As we explained in the earlier sections, 
the phonons are shifted in frequency and their degeneracy lifted by the interaction with the optical magnon [Fig.~\ref{fig:dielectric}(c-d)], and the effect is strongest for modes that lie close in the spectrum.
In turn, the magnon hybridizes with the phonon (see Sec.~\ref{sec:lattdyn}), and thereby acquires a significant dipolar strength. As a consequence, the intensity of the magnon-like peak at 28.43 meV [panel (d)] is now more than three orders of magnitude larger than the original bare magnon feature [panel (b)].
The amplified intensity of the renormalized optical magnon resonance, arising from the spectral weight transfer from the nearby $E^+_u$ phonon mode, is consistent with experimental observations in other material classes,~\cite{ValdesPRB07,pimenov2006,Mashkovich21} and suggests that such feature should be readily detectable through optical methods.

\blue{As we noted earlier, however, our first-principles model yields an OM that is approximately 8 meV higher in energy than the experimental value~\cite{ChenPRX18}.
Such a discrepancy likely originates from the LDA overestimation of the exchange couplings that contribute to
the static magnetic Hessian $K$~\cite{Gorni_bulk_PRB23,delugas_mono_PRB23}.
This is not \emph{per se} a big issue, as long as the representation of the physics is 
qualitatively correct.
Unfortunately, in our case the calculated dipolar strengths and frequency splittings strongly depend on the relative
position of the magnon and phonon resonances.
Since phonon frequencies are remarkably accurate within DFPT, the misalignment of the magnon mode can have
direct consequences on the predicted spectra.}

\begin{figure}
 \centering
 \includegraphics[width=1.0\columnwidth]{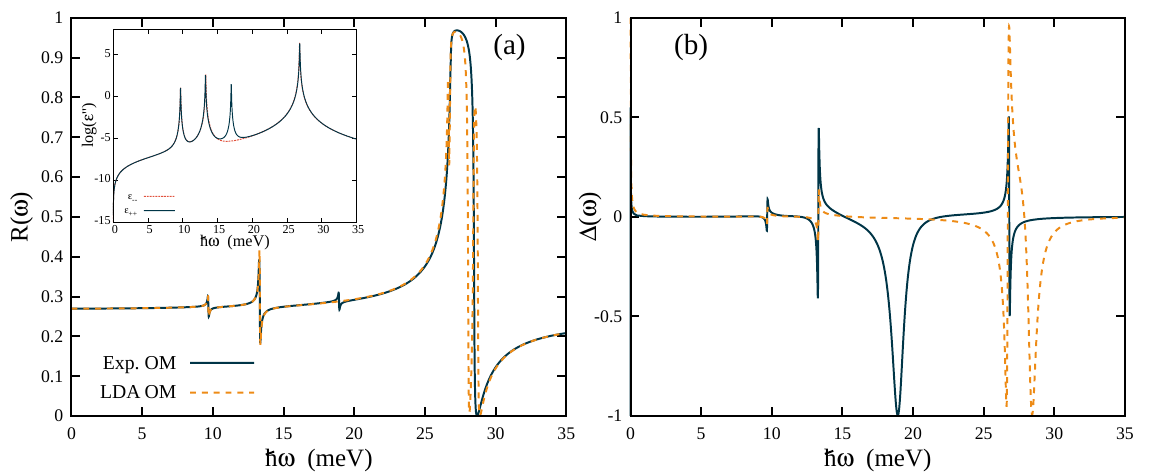}
 \caption{(a) Reflectivity spectrum for incidence normal to the $x$--$y$ plane and (b) magnetic circular dichroism signal of CrI$_3$. The results are obtained from the full LDA dielectric tensor (orange, dashed) and from an effective dielectric tensor calculated after adjusting the coefficients of the magnetic Hessian $K$ to reproduce the experimental optical-magnon frequency (blue, solid). The inset shows the logarithmic imaginary part of the dielectric tensor computed with the adjusted Hessian.}
 \label{fig:reflectivity}
\end{figure}

\blue{To obtain a more realistic picture of the experimental situation, in the following we perform a computational experiment where
we modify the $K$ coefficient of the OM to match the experimental frequency of 19 meV~\cite{ChenPRX18}, while leaving the remainder parameters unaltered.
We then recalculate the absorptive part of the dielectric tensor via Eqs.~\eqref{eq:E2_BtoH} applied to the revised parameter set;
the result is shown in the inset of Fig.~\ref{fig:reflectivity}(a).
As expected, the energy splitting of the chiral $E_u$ phonon at 28 meV is smaller in magnitude ($\Delta \w =$0.032 meV) and opposite in
sign compared to our raw first-principles data of Fig.~\ref{fig:dielectric}(d).
The sign of the effect, in particular, is now consistent with the predictions of Ref.~\cite{RenPRX24}, where the energy ordering
of the phonon and magnon resonances matches that of our modified model. 
Interestingly, the optical magnon now lies much closer in energy to the $E_u$ doublet at 13.3 meV. As a consequences, the latter mode displays
a sizeable frequency splitting ($\Delta \w =$$-$0.026 meV) as well, and significantly contributes to the dipolar strength of the magnon.
}

To further illustrate these points, 
we calculate the reflectivity spectrum of optical waves normal to the $x$--$y$ plane from the Cartesian RI dielectric tensor, 
\begin{equation}
R(\omega)=\left| \frac{\sqrt{\varepsilon^{\rm RI}_{\bf q}(\omega)}-1}{\sqrt{\varepsilon^{\rm RI}_{\bf q}(\omega)}+1} \right|^2,
\end{equation}
where $\varepsilon^{\rm RI}_{\bf q}$ represents the projection of the dielectric tensor on an in-plane direction $\bf q$ of the crystal.
The result is shown in Fig.~\ref{fig:reflectivity} (a) for the relevant frequency range.
As observed experimentally~\cite{TomarchioSR21,TomarchioM23}, the calculated reflectivity spectrum of CrI$_3$ is dominated 
by a broad, intense signal that spans frequencies between 26 and 29 meV, corresponding to light absorption by the highest-energy $E_u$ phonon mode.
\blue{In addition, within our raw first-principles model we can identify a distinct, intense peak around 28.5 meV, which is due to the optical magnon resonance.
Our modified model, on the other hand, 
exhibits much weaker magnon-related features, with only a small signal at the OM resonance.
Still, the optical magnon feature is comparable in magnitude to that of the IR-active phonon at 13.3 meV,
which suggests the concrete possibility of detecting it in reflectivity spectra.}

Finally, we evaluate the magnetic circular dichroism (MCD) spectrum arising from the interacting magnon-phonon system.
The MCD signal is calculated as,~\cite{goldbook}
\begin{equation}
 \Delta(\omega)=\frac{\varepsilon_{--}^{\rm RI,A}(\omega) - \varepsilon_{++}^{\rm RI,A}(\omega)}{\varepsilon_{--}^{\rm RI,A}(\omega) + \varepsilon_{++}^{\rm RI,A}(\omega)},
\end{equation}
where the superscript ``A'' denotes the antihermitian (dissipative) part of the dielectric tensor.
\blue{The raw LDA results and those based on the modified model discussed in the previous paragraph are presented in Fig.~\ref{fig:reflectivity}(b).}
\blue{In both cases, a broad negative peak appears due to absorption by the optical magnon, accompanied by three phonon-related features. The amplitudes of the latter are clearly enhanced through coupling with the OM, and their signs depend on the relative position in energy with respect to the magnon.}
\blue{All in all,
our results emphasize the importance of treating magnons and phonons self-consistently in the calculation of the frequency-dependent dielectric function, as most of the aforementioned features 
cannot be captured within approximations where the spin-wave dynamics is not fully accounted for.~\cite{CalandraPRB10,FiorazzoPRB25}}

\subsection{Electromagnons}

The analysis of the previous section has revealed an interesting interplay of spin and phonon modes in determining the infrared spectrum of the crystal.
In this Section we clarify how the magnetic degrees of freedom respond to the external electric field $\bm{\mathcal{E}}$,
giving rise to a hybrid spin-phonon electromagnon feature.

\blue{The steady-state oscillation amplitude of a given spin mode is readily given in terms of the second derivatives of either $F(\tau,H)$
or $\mathcal{F}(f,H)$, depending on whether the ions are clamped ($F$) or allowed to move ($\mathcal{F}$). 
Straightforward application of Eq.~\eqref{mlam1} yields
\begin{align}
\label{eq:emag_ci}
 \frac{\partial m_j}{\partial \mathcal{E}_\alpha}\Big|_{F}= &
-[U^{(ss)}]^{-1}_{jl} U_{l \mathcal{E}_\alpha} = -\chi_{jl} \frac{\partial H_l}{\partial \mathcal{E}_\alpha}\Big|_{U}, 
\end{align}
for the clamped-ion response, and 
\begin{align}
\label{eq:emag_ri}
\frac{\partial m_j}{\partial \mathcal{E}_\alpha}\Big|_{\mathcal{F}}= &
-U^{-1}_{ja} U_{a \mathcal{E}_\alpha}  \nonumber \\ 
=& - \chi^{\rm RI}_{jl} \left[ \frac{\partial H_l}{\partial \mathcal{E}_\alpha}\Big|_{U} 
 +  \frac{\partial H_l}{\partial \tau }\Big|_{U} [U^{(pp)}]^{-1}_{\tau\tau'} [Z_{\alpha,\tau'}]^* \right]
\end{align}
for its relaxed-ion counterpart. In the second line of Eq.~\eqref{eq:emag_ri} we have used 
$U^{-1}_{j \tau'} = -\chi^{\rm RI}_{jl} U^{(sp)}_{l\tau} [U^{(pp)}]^{-1}_{\tau\tau'}$,
and observed that $U_{\tau \mathcal{E}_\alpha}=[Z_{\alpha,\tau}]^*$ (the absorptive non-Hermitian part can be safely neglected in the mixed derivatives at frozen spins).
This way, Eq.~\eqref{eq:emag_ci} and Eq.~\eqref{eq:emag_ri} are written in the exact same form as minus the spin susceptibility times the first-order Zeeman fields;
the difference lies in that both quantities are calculated at the CI level
in Eq.~\eqref{eq:emag_ci}, at the RI level in Eq.~\eqref{eq:emag_ri}.
In the following we discuss our numerical results by adopting, as earlier in this Section, a circularly polarized basis for the optical magnon and applied electric field.}

\begin{figure}
 \centering
 \includegraphics[width=1.0\columnwidth]{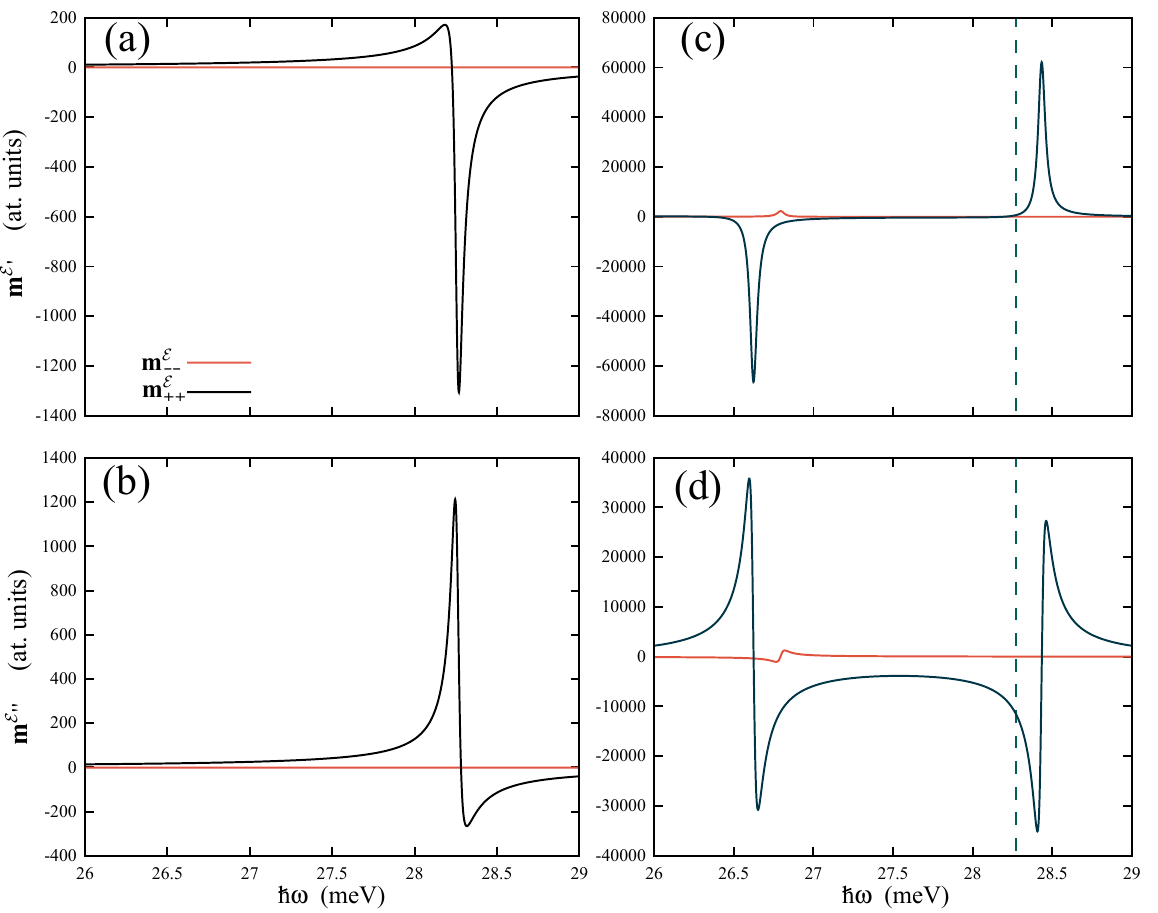}
 \caption{Real (top panels) and imaginary (bottom panels) parts of the diagonal components of the local
 magnetoelectric susceptibility, expressed in the circularly polarized basis and for the optical magnon, as a function of the frequency.
 Left (right) panels show
 clamped-ion (relaxed-ion) flavour of the susceptibilities.
 The vertical grey-dashed line depicts the position of the CI optical magnon resonance. }
 \label{fig:magnetoelectric}
\end{figure}

Our CI results are shown in Fig.~\ref{fig:magnetoelectric}(a-b).
As in the case of the dielectric tensor, the $++$ and $--$ components are degenerate in the static limit, and progressively split as the frequency increases, until the $++$ component diverges at the bare OM frequency.
Interestingly, the real and imaginary parts of the response differ significantly in shape from those seen in the dielectric tensor [Fig.~\ref{fig:dielectric}], and indicate a nontrivial phase shift for the OM response to the field.
To understand such phase shift, one can write the CI spin response to the field as a scalar function of frequency,
\begin{equation}
\frac{\partial m_+}{\partial \mathcal{E}_+} = 
\frac{H^\mathcal{E}_+}{K\mp \w G - \w^2 M}
\end{equation}
The calculated phases of the coupling coefficients $H^\mathcal{E}_\pm$ in the static limit are approximately $\mp 126^\circ$ in the $\pm \pm$ sectors, and evolve almost linearly with frequency, reaching $-130^\circ$ and $123^\circ$ at 30 meV. 
This unusual phase mismatch is due to the out-of-plane stacking in bulk CrI$_3$, which lowers the symmetry group of
the individual layers; indeed, in the monolayer limit an exact $90^\circ$ phase shift is expected.

As shown in Figs.~\ref{fig:magnetoelectric} (c) and (d), the result changes drastically when both spin and lattice degrees of freedom are allowed to evolve in response to the field. 
Compared to the CI case, we observe the following: (i) the magnitude of the $++$ response increases by two orders of magnitude, (ii) the phase shift changes to a nearly constant value of $-97^\circ$ across the entire frequency window, and (iii) the OM-like and the $E_u^+$ phonon-like resonances exhibit similar spectral weights in terms of magnitude, but with opposite signs.
All these features can be explained by taking a closer look at Eq.~\eqref{eq:emag_ri}. First, in a vicinity of the $E_u$ resonance one can expect a large enhancement in the magnitude of the response because of the divergent nature of the bare phonon propagator (second term in the square bracket). Next, given that the lattice-mediated response dominates over the direct one (first term in the square bracket), the overall phase delay is now dictated by the spin-phonon coupling coefficient, $H^\tau$, instead of $H^\mathcal{E}$. 
Finally, note that the divergency in the bare phonon propagator cancels out exactly with a zero of $\chi^{\rm RI}$, yielding an equal and opposite response at the two hybrid spin-phonon resonances.
Thus, the above can be regarded as general features of hybrid electromagnons where the interaction with the electric field is mediated by the phonon dipolar strength; a more quantitative derivation to support this interpretation is provided in Appendix~\ref{sec:model}.

\blue{
\section{Antiferromagnetic C\lowercase{r}$_2$O$_3$ \label{sec:Cr2O3}}}

\blue{Cr$_2$O$_3$ is an antiferromagnetic insulator with a corundum structure containing two formula units per primitive cell (Fig.~\ref{fig:Cr2O3_structure}).
It was the first crystal in which the linear magnetoelectric effect was experimentally observed~\cite{astrovJETP60,FolenPRL61} following Dzyaloshinskii's prediction~\cite{DzyaloshinskiJETP60}, and is now regarded as the prototypical local-moment magnetoelectric.
Despite extensive study, the role of collective spin-wave excitations in its optical and magnetoelectric properties remains only partially understood.}

\blue{For instance, recent THz pump-probe experiments by Bilkyl \emph{et al.}~\cite{bilyk2025l} showed that the antiferromagnetic mode (acoustic magnon) of Cr$_2$O$_3$ can be driven not only by the Zeeman component of a THz pulse, as expected by the established selection rules, but also by its electric field.
Remarkably, the dynamics induced by the two channels have comparable amplitudes and scale linearly with field strength.
This was interpreted as a dynamical magnetoelectric effect, though the microscopic mechanism --such as possible coupling to vibrational modes-- remains unresolved.}

\blue{In the following, we analyze the electromagnetic susceptibilities of Cr$_2$O$_3$ across the vibrational spectrum, with particular attention to a magnon-phonon spectral-weight transfer that proves essential for quantitatively accounting for the THz response observed by Bilkyl \emph{et al.}~\cite{bilyk2025l}.
To establish a clear framework, we first introduce an adiabatic model of antiferromagnetic magnons formulated in terms of static collective magnetic modes.
This construction provides an intuitive basis for interpreting the magnonic features that emerge in the macroscopic response functions.}

\blue{
\subsection{Static magnetic susceptibility \label{sec:Cr2O3_static_chi}}}

\blue{In Cr$_2$O$_3$, the magnetic degrees of freedom span an eight-dimensional space corresponding to the in-plane components of the four Cr magnetic moments (or local Zeeman fields).
The inverse local susceptibility, $\boldsymbol{\chi}^{-1}={\bf U}^{(ss)}$, is therefore an $8\times8$ tensor with four doubly-degenerate eigenvalues.
To start with, we represent it 
onto the following Cartesian basis of collective spin cantings, 
\begin{equation}
\label{eq:cr2o3modes}
\mathbf{e}=\frac{1}{2} \,\,
\begin{blockarray}{*{4}{c} l}
\begin{block}{*{4}{>{$\footnotesize}c<{$}} l}
  Cr$_1$ & Cr$_2$ & Cr$_3$ & Cr$_4$ & \\
\end{block}
\begin{block}{[*{4}{r}]>{$\footnotesize}l<{$}}
    1 & -1 & 1 & -1 \quad & \, ${\bf e}_1$ \\
    1 & -1 & -1 & 1 \quad & \, ${\bf e}_2$ \\
    1 & 1 & 1 & 1   \quad & \, ${\bf e}_3$  \\
    1 & 1 & -1 & -1 \quad & \, ${\bf e}_4$\\
\end{block}
\end{blockarray}
,
\end{equation}
where the individual entries are $2\times2$ identity matrices in the Cartesian space of the 
in-plane magnetic moments.
This basis is very convenient, as it can be directly linked to key results established in
the existing literature on antiferromagnets~\cite{KittelPR52,Keffer-53,MuPRM19}:
${\bf e}_1$, of $E_u$ character, 
corresponds to the primary (antiferromagnetic) order parameter, often referred to as ${\bf L}$~\cite{MuPRM19};
${\bf e}_3$ (of $E_g$ character) corresponds to the ferromagnetic mode ${\bf M}$ 
(not to be confused with the mass tensor, which we indicate with the same symbol),
with all spins canting in the same in-plane direction;
${\bf e}_2$ ($E_g$) and ${\bf e}_4$ ($E_u$) are secondary antiferromagnetic modes, 
indicated as ${\bf L}''$ and ${\bf L}'$ in Ref.~\onlinecite{MuPRM19}, respectively.
Note that the ${\bf e}_j$ mode amplitudes are expressed as canting angles: they are linearly related
to the aforementioned order parameters by a constant prefactor, see Section~\ref{sec:me}.}

\blue{Within this representation, the 
static Hessian $\mathbf{K}={\bf U}^{(ss)}(\w=0)$ adopts the following structure,
\begin{equation}
\label{eq:K_modes}
{\bf K} = \begin{pmatrix} 
K_1  & 0 & 0 & D_u \\
0 & K_2  & D_g & 0 \\
0 & D_g^{T} & K_3  & 0 \\
D_u^T & 0 & 0 & K_4 
\end{pmatrix},
\end{equation}
where the individual elements are $2\times 2$ matrices 
in the form
$K_j = k_j \mathcal{S}$ and $D_{u,g} = d_{u,g} \mathcal{A}$, and
\begin{equation}
\mathcal{S} = \begin{pmatrix} 1 & 0 \\
 0 & 1 
 \end{pmatrix},\qquad  \mathcal{A} = \begin{pmatrix} 0 & -1 \\
 1 & 0 
 \end{pmatrix}.
 \end{equation}
The calculated values are reported in Table~\ref{tab:afm_adi_param}.
$k_1$, related to the (weak) magnetocrystalline anisotropy, is by far the smallest parameter consistent with its relativistic origin. 
(Incidentally, our Sternheimer-based implementation reproduces the 
Goldstone character of ${\bf e}_1$ within machine precision when SOC is switched off.
This represents a clear advantage over approaches relying on truncated sums over excited states, where enforcing 
the Goldstone rule has proven particularly challenging in antiferromagnets~\cite{SkovhusPRB22}.)}

\begin{table}
\setlength{\tabcolsep}{6pt}
\begin{center}
\caption{Adiabatic expansion of the matrix ${\bf U}^{(ss)}$, obtained by fitting a fourth-order polynomial in $\omega$. The Hessian (${\bf K}$) and mass (${\bf M}$) tensors follow the structure given in Eq.~\eqref{eq:K_modes}, while the Berry curvature tensor (${\bf G}$) is represented according to Eq.~\eqref{eq:G_modes}. The two bottom rows show the off-diagonal entries of ${\bf K}$ and ${\bf M}$}
\begin{tabular}{rrrr}\hline\hline
   \T\B $i$ & \multicolumn{1}{c}{$k_{i}$ (meV)} & \multicolumn{1}{c}{$g_{i}$} & \multicolumn{1}{c}{$m_i$ 
     } (eV$^{-1}$) \\ \hline
   \T\B 1 &   0.060  & 1.3109    & 0.081 \\
   \T\B 2 & 168.433  & 1.3813    & 0.154 \\
   \T\B 3 & 217.315  & 0.0002    & 0.200 \\
   \T\B 4 & 267.040  & $-$0.0013 & 0.223 \\
   \hline \hline
   \T\B $p$ & \multicolumn{1}{c}{$d^{\bf K}_{p}$ (meV)} & &   \multicolumn{1}{c} {$d^{\bf M}_{p}$ (eV$^{-1}$)}\\ \hline
   \T\B $g$ &    2.135 &  & 0.0002\\
   \T\B $u$ & $-$1.869 &  & 0.0002\\
   \hline \hline
\end{tabular}
\label{tab:afm_adi_param}
\end{center}
\end{table}

\blue{The remainder diagonal elements $k_j$ are more than three orders of magnitude larger; of them, $k_3$ is the most interesting physically
as it directly relates to the static macroscopic magnetic susceptibility.
The two off-diagonal parameters are intermediate in magnitude, and have to do with the two 
energy invariants derived from the Dzyaloshinskii-Moriya interaction (DMI), of the type $d_u (L'_x L_y-L'_y L_x)$ and $d_g (L''_x M_y-L''_y M_x)$~\cite{MuPRM19}.
(Both $d_u$ and $d_g$ are of the order of 2 meV in our calculations, consistent with the results of
Ref.~\onlinecite{MuPRM19} for the DMI parameter $D$.)
As a consequence of the latter, the eigenvectors of the static spin susceptibility matrix $\bm{\chi}(0)$ differ from the
basis vectors ${\bf e}_j$ for a slight intermixing between modes of the same parity. The angles of rotation in the
${\bf e}_{1,4}$ and ${\bf e}_{2,3}$ subspaces are, respectively, $\alpha_u\sim0.4^{\circ}$ and $\alpha_g\sim2.5^{\circ}$.
These are very small effects, but enough to produce a significant decrease of the anisotropy parameter,
from the ``collinear''~\cite{MuPRM19} value $k_1$ to $\tilde{k}_1=0.047$ meV.}

\begin{figure}
 \centering
 \includegraphics[width=0.8\columnwidth]{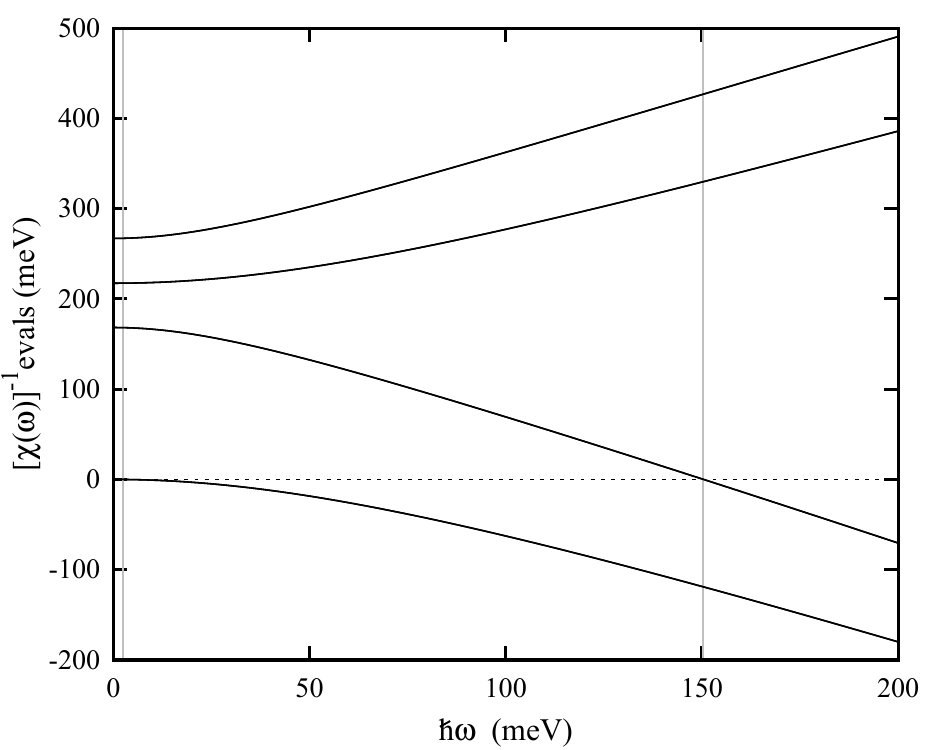}
 \caption{Eigenvalues of the inverse local spin susceptibility of Cr$_2$O$_3$ as a function of frequency. Vertical grey lines signal the magnon resonance frequencies.}
 \label{fig:evals}
\end{figure}

\blue{
\subsection{Antiferromagnetic magnons}}

\blue{The bare magnon resonances are defined, within our formalism, by
the frequencies at which ${\bf U}^{(ss)}(\w)$ becomes singular.
In Fig.~\ref{fig:evals} we show the frequency dispersion of the (doubly degenerate) eigenvalues of 
${\bf U}^{(ss)}(\w)$, where the two zeros are indeed evident.  
To understand the physics underlying these spectral curves, 
in the following we complement the discussion of the Hessian (see previous section) with an analysis
of the Berry curvature tensor, ${\bf G}$.
(For simplicity we restrict our analysis to the FOA level of theory for the time being, where
the inverse susceptibility can be conveniently represented as ${\bf U}^{(ss)} \simeq {\bf K} + i\w {\bf G}$, see Section~\ref{sec:linearw}).}
\blue{Its representation on the basis of Eq.~\eqref{eq:cr2o3modes} reads as
\begin{equation}
\label{eq:G_modes}
{\bf G} = \begin{pmatrix} 
0 & G_3 & G_1 & 0 \\
-G_3 & 0 & 0 & G_2 \\
G_1 & 0 & 0 & G_4 \\
0 & G_2 & -G_4 & 0 
\end{pmatrix},
\end{equation}
where the individual $2\times 2$ blocks are in the form
$G_{1,2} = g_{1,2} 
\mathcal{A}$  and $G_{3,4} = g_{3,4} \mathcal{S}$.
Due to the time-odd nature of the Berry curvature, in conjunction with the underlying $PT$ symmetry
of the Cr$_2$O$_3$ crystal, the nonzero blocks of ${\bf G}$ only occur
between irreps of different parity, which are 
therefore allowed to mix at finite frequency~\cite{RenPRX24}.
Such mixing primarily occurs between ${\bf e}_1-{\bf e}_3$
and ${\bf e}_2-{\bf e}_4$ pairs, while the remaining symmetry-allowed couplings (${\bf e}_1-{\bf e}_2$ and
${\bf e}_3-{\bf e}_4$) are negligibly small (see Table~\ref{tab:afm_adi_param}).
The peculiar frequency dispersion of the susceptibility spectrum (Fig.~\ref{fig:evals}) is precisely due to the
(1,3) and (2,4) matrix elements of ${\bf G}$.}

\blue{The physical reason why the Berry curvature strongly couples (1,3) and (2,4) and not other pairs
can be understood by recalling the classic Kittel's arguments about antiferromagnetic resonances~\cite{KittelPR52}.
Suppose we neglect for a moment the small DMI parameters $d_{u,s}$, which are largely irrelevant
to the present qualitative discussion.
Then, by solving the adiabatic equation of motion ${\rm det}({\bf K} + i\w{\bf G})=0$, we obtain a doubly degenerate AM mode in the form ${\bf e}^{\rm AM} =( \alpha e_{1x} + \beta e_{3y}, \alpha  e_{1y} - \beta e_{3x})$, where the resonance frequency and aspect ratio are given by
\begin{equation}
\label{ellw}
\frac{\alpha}{\beta} \simeq \sqrt{\frac{k_3}{k_1}}, \qquad \w_{\rm AM} \simeq \frac{\sqrt{k_1 k_3}}{|g_{1}|}.
\end{equation}
(Similar formulas hold for the OM, which is associated with the ${\bf e}_2 - {\bf e}_4$ pair.)
Thus, magnons in Cr$_2$O$_3$ naturally emerge from our formalism as \emph{elliptically polarized} modes~\cite{KittelPR52}, where the two orthogonal components have different amplitude and opposite parity.}
\blue{This behavior reflects a universal characteristic of easy-axis antiferromagnets in zero field:
the natural precession of the individual spins results in the in-plane components of ``up'' and ``down'' 
sublattices rotating in opposite directions~\cite{KittelPR52}.
This implies that, in the course of an oscillation cycle, the spin canting pattern will visit two 
different modes that are related by a sign switch on the Cr$_2$ and Cr$_4$ sites; these are precisely
\blue{(1,3) and (2,4)}.}
\blue{Moreover, by incorporating into this reasoning the mode mixing associated with the $d_{u,s}$  parameters, one recovers the full semiclassical picture~\cite{Keffer-53}, in which the two spin sublattices precess with distinct aspect ratios.}

\begin{table}[!b]
\setlength{\tabcolsep}{6pt}
\begin{center}
\caption{\label{tab:afm_magnons} Magnon resonance frequencies (in meV), determined from Eq.~\eqref{ellw}; from the solution of the adiabatic equation of motion, excluding (FOA) and including (SOA) the mass contributions; and from the zeros of the curves in Fig.~\ref{fig:evals} (TD-DFPT).}

\begin{tabular}{c|cccc}\hline\hline
   \T\B & Eq.~\eqref{ellw} & FOA & SOA & TD-DFPT  \\
   \hline
   \T\B AM & 2.754 & 2.445 & 2.432 & 2.431\\
   \T\B OM & 153.537 & 153.498 & 150.447 & 150.438 \\
   \hline \hline
\end{tabular}
\end{center}
\end{table}

\blue{We report in Table~\ref{tab:afm_magnons} the numerical results for the magnon frequencies calculated
at various levels of theory: Eq.~\eqref{ellw}, FOA, SOA and full TD--DFPT.
Overall, our first-principles implementation reproduces the intuitive physical picture outlined above with remarkable 
accuracy.
The FOA values are in excellent agreement with Eq.~\eqref{ellw}: the latter only differ from the former by the neglect of the
off-diagonal elements ($d_{u,g}$) of the Hessian, whose most significant consequence is the renormalization of the anisotropy 
parameter $k_1$.
The main change occurs when moving from the FOA to the SOA, where inclusion of the effective magnon masses results in 
a red shift of the optical magnon frequency by approximately 3 meV.
(The time-even mass tensor ${\bf M}$ bears an analogous structure as the Hessian ${\bf K}$; the calculated values of the
independent entries are also reported in Table~\ref{tab:afm_adi_param}.)
Such a correction can be understood as a renormalization of the Berry curvature parameter $g_2$ 
by a factor of $\gamma \simeq 1+(m_2k_4 + m_4 k_2)/(2g_2^2)$: by using the values of Table~\ref{tab:afm_adi_param} we obtain $\gamma=1.02$,
consistent with the calculated $\approx 2\%$ redshift.
This result highlights the necessity of accounting for the magnon inertia in developing quantitatively accurate adiabatic models of spin dynamics.
As in CrI$_3$, the SOA model matches the full TD--DFPT values essentially to machine precision.}

\begin{figure*}
 \centering
 \includegraphics[width=1.0\textwidth]{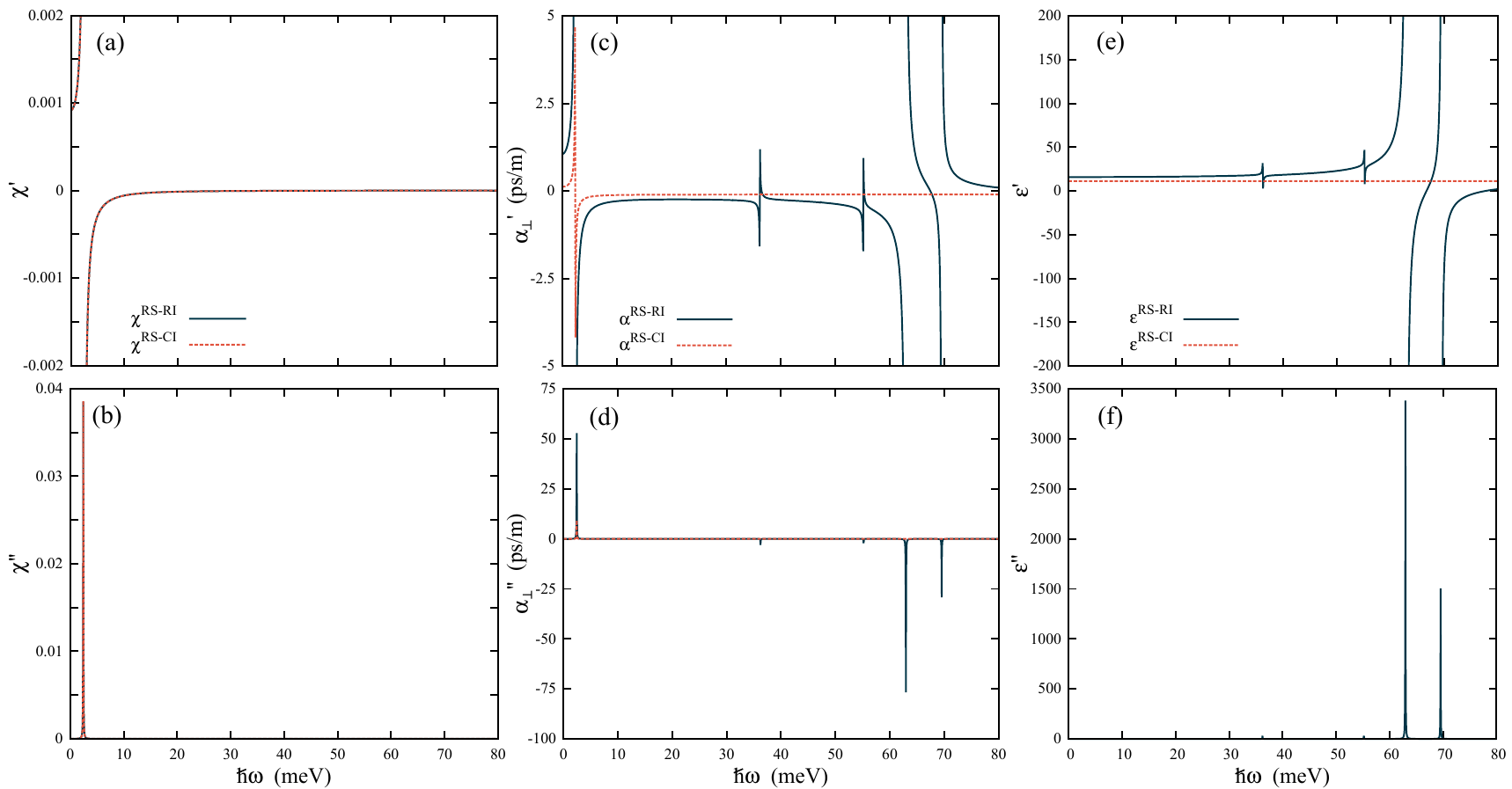}
 \caption{Frequency dispersion of the three electromagnetic linear-response tensors in Cr$_2$O$_3$: (a,b) magnetic susceptibility, (c,d) magnetoelectric tensor, and (e,f) dielectric tensor. Both the clamped-ion (red dashed) and relaxed-ion (dark-blue solid) cases are shown, with spins  relaxed in all calculations. For clarity, only the $++$ in-plane component is plotted; the $--$ component is identical by symmetry.}
 \label{fig:Cr2O3_sus}
\end{figure*}

\blue{To conclude this section, we note that our LDA treatment of exchange and correlation effects yields magnon resonances that
do not reproduce the experimental values (0.68 and $\sim55$ meV~\cite{SamuelsenPhys70}) very accurately. 
Regarding the acoustic magnon, the discrepancy can be rationalized by looking at the
dependence of the Hessian ${\bf K}$ on the geometry of the simulation cell.
Our calculations on the experimental Cr$_2$O$_3$ structure yield an unstable
spin Hessian with a negative anisotropy parameter of $k_1=-0.051$ meV. This
points to the well-known qualitative failure of LDA (and Hubbard-corrected LDA+U~\cite{MuPRM19}) at reproducing the
correct easy-axis anisotropy in this material.
The relaxed structure within LDA, on the other hand, yields a severe underestimation of the rhombohedral angle, 
significantly worsening (we find $k_1=-0.135$ meV) the issue~\cite{MoseyPRB07}.
Consistent with earlier works~\cite{MuPRM19,RenPRX24}, we find that a tensile in-plane strain of 2\% compared to the
experimental lattice parameter is enough to restore the observed easy axis~\footnote{It would have certainly been possible to achieve a better match by using a slightly smaller strain; however,
  since this detail is largely irrelevant for the discussion that follows, we have refrained from attempting further adjustments.}.
  }

\blue{In the case of the OM, our calculated resonance frequency is strongly overestimated by almost a factor of 3
compared to the experiment. Note that this is not an artefact of the artificially strained structure, since we
find essentially the same value ($\w_{\rm OM} = 154.4$ meV) in the experimental structure.
As we have pointed out in the case of CrI$_3$, local and semilocal approximations to DFT suffer~\cite{Gorni_bulk_PRB23,delugas_mono_PRB23} from
a systematic overestimation of magnetic exchange interactions, likely due to the well-known tendency of LDA to
delocalize the atomic orbitals.
We believe that similar detrimental effects are responsible for the large overshoot of $\w_{\rm OM}$ in Cr$_2$O$_3$.
For Cr$_2$O$_3$, previous studies have shown that including a Hubbard $U$ correction softens the bare LDA magnons~\cite{SkovhusPRB22,RenPRX24}, but at the cost of introducing both qualitative and quantitative anomalies in their momentum dispersion~\cite{SkovhusPRB22}.
Since the aforementioned disagreement bears limited consequences on the discussion that follows,
we have not explored such a route in this work. Indeed, the OM plays essentially no role in the coupling of Cr2O3 to external fields, 
which are dominated by the AM in the low-frequency region of the spectrum.} 

\blue{
\subsection{Macroscopic electromagnetic susceptibilities}}

\blue{In Fig.~\ref{fig:Cr2O3_sus}, we present the in-plane components of the relaxed-spin
tensors that describe the linear electromagnetic response of Cr$_2$O$_3$ across 
the vibrational frequency range.
We consider, in particular, the dielectric constant $\e$, the
macroscopic magnetic susceptibility $\chi$ and the magnetoelectric coefficient $\alpha$.
Their numerical values are calculated following the recipes of Sec.~\ref{sec:sus}, i.e., as second derivatives
of the appropriate enthalpy functional with respect to macroscopic electric or Zeeman fields.
(We use $-F_{\lambda\lambda'}/\Omega$ or $\mathcal{F}_{\lambda\lambda'}/\Omega$ within 
clamped-ion or relaxed-ion conditions, respectively.)
}

\blue{For the magnetic susceptibility, we obtain in the static limit a value that is essentially insensitive to ionic relaxations, $\chi^{\rm CI}\approx\chi^{\rm RI}=9.19\times10^{-4}$.
Compared with previous LDA+U results~\cite{BousquetPRL11}, our value is smaller by roughly a factor of two.
This discrepancy is consistent with the LDA overestimation of the magnetic-Hessian parameters discussed above [Eq.~\eqref{eq:K_modes}]: the susceptibility is largely controlled by the inverse of $k_3$, which in our calculation is 217.3 meV, whereas Ref.~\onlinecite{RenPRX24} reports a substantially smaller value of 89.9 meV within the LDA+U framework.
The low-frequency part of the spectrum is dominated by the acoustic magnon resonance, consistent with the 
discussion of the previous subsection. (The AM resonance is a coupled oscillation of the ferromagnetic and antiferromagnetic order parameters.)
In contrast, the optical magnon contributes negligible spectral weight,
which is in line with its antiferromagnetic character.
The contribution of the lattice modes to the magnetic response is even smaller, indicating that the coupling between
phonons and the spin mode ${\bf e}_3$ is weak on the scale of Fig.~\ref{fig:Cr2O3_sus}(a).
And indeed, the frequency shifts that are associated with magnon-phonon interactions are generally small:
For example, the acoustic magnon resonance only shifts by $-0.17$ $\mu$eV between CI and RI conditions,
and its spectral weight remains essentially unchanged.}

\blue{The dielectric tensor shows the opposite trend, with the vibrational resonances outweighing
the magnetic ones by five orders of magnitude.
Their respective contribution (28.7\% and 0.002\%) to the static dielectric constant reflects this behavior,
pointing to a negligible dipolar strength of the spin modes compared to the phonons.
The most prominent features correspond to the two $E$ modes at 62.9 and 69.5 meV, while other lattice modes (e.g.,
those at 36.2 and 55.2 meV) display a comparatively much weaker signal.}

\blue{Finally, the magnetoelectric tensor exhibits an intermediate picture, with the acoustic magnon and IR-active
phonons yielding comparable spectral weight.
Interestingly, the sign of the magnetoelectric response appears to be reversed at the phonon resonances compared to the
magnon peak. This is is reminiscent of our discussion of the electromagnon in CrI$_3$, although 
the mechanism leading to such a behavior appears to be different (see next subsection~\ref{sec:me}).
Ionic relaxations also lead to a marked increase in the spectral weight of the acoustic magnon contribution to the magnetoelectric tensor.
Projection of the lattice-mediated contribution reveals that most of this weight is transferred by the $E$ mode at $\sim 62.9$ meV.
A similar trend is observed in the static limit: the magnetoelectric response is enhanced by nearly an order of magnitude due to the same $E$ mode, with $\alpha^{\rm CI}_{\perp}=0.12$ and $\alpha^{\rm RI}_{\perp}=1.05$ ps/m.
Our result for thelattice-mediated contribution,  $\alpha^{\rm LM}_{\perp}=\alpha^{\rm RI}_{\perp}-\alpha^{\rm CI}_{\perp}$, is in good agreement with previous reports~\cite{IniguezPRL8,BousquetPRL11,MalashevichPRB12,YePRB14,BousquetJPCM24}.
$\alpha^{\rm CI}_{\perp}$, on the other hand, appears somewhat underestimated compared to literature values~\cite{BousquetPRL11,MalashevichPRB12,YePRB14,BousquetJPCM24}.
We ascribe this discrepancy to the epitaxial strain applied to our simulation cell: indeed, when using the unstrained experimental structure, we obtain $\alpha^{\rm CI}_{\perp}=0.36$ (and $\alpha^{\rm RI}_{\perp}=1.14$ ps/m), in closer agreement with Refs.~\onlinecite{BousquetPRL11,MalashevichPRB12}.
This sensitivity, which is only weakly reflected in $\alpha^{\rm LM}_{\perp}$ and in the magnetic susceptibility, originates from 
an unusually strong strain dependence of the coupling parameter $k^{\mathcal{E}}_3$ (see next subsection).}

\blue{The large spectral-weight transfer observed for the acoustic magnon is directly relevant to the recent experiments by Bilkyl \emph{et al.}~\cite{bilyk2025l},
reporting on the resonant excitation of the acoustic magnon in Cr$_2$O$_3$ via a THz electric-field pump. 
To trace a closer link to Ref.~\cite{bilyk2025l}, and more generally to assess the respective role of the spin and phonon degrees of freedom in the dynamical magnetoelectric response of
Cr$_2$O$_3$, we provide a more detailed analysis in the following.}

\blue{
\subsection{Dynamical magnetoelectric coupling \label{sec:me}}}

\blue{
First, we analyze how the constitutive magnetic modes of Eq.~\eqref{eq:cr2o3modes} couple to the external electric field. 
Based on Eq.~\eqref{eq:emag_ri}, the Zeeman torques induced by an electric field $\mathcal{E}_\beta$ at frequency $\w$ on a magnetic mode of Eq.~\eqref{eq:cr2o3modes} --characterized by a mode index $n$ and Cartesian direction $\alpha$-- can be written as
\begin{equation}
\label{lamb}
\begin{split}
\frac{\partial H_{n\alpha} }{\partial \mathcal{E}_\beta} \Big|_{\mathcal{U}} 
 =& \frac{\partial H_{n\alpha}}{\partial \mathcal{E}_\beta} \Big|_{U} + \sum_\nu \frac{ \partial H_{n\alpha} }{\partial u_{\nu\gamma}}\Big|_{U}
   \frac{ \partial u_{\nu\gamma} }{\partial \mathcal{E}_{\beta}}\Big|_{\mathcal{U}}
  \end{split}
\end{equation}
where the summation on the rhs is performed over the $E$ doublets ($\nu$ labels the irrep, $\gamma=x,y$
discriminates the two Cartesian components and $u_{\nu \gamma}$ is the mode amplitude), and the frozen-spin
functionals $U$ and $\mathcal{U}$ refer to clamped-ion and relaxed-ion conditions, respectively.
The first term in Eq.~\eqref{lamb} describes a direct coupling, which is of purely electronic origin;
the second term embodies the lattice-mediated contribution of each phonon irrep, and 
corresponds to the second term in the square brackets of Eq.~\eqref{eq:emag_ri}.
To simplify the analysis, we shall neglect the dependence of $Z_{\beta,\nu\gamma}$ and of the force-constant matrix on frequency henceforth;
this is justified since both appear to have negligible impact on the results.
We then arrive at
\begin{equation}
 \frac{ \partial u_{\nu\gamma} }{\partial \mathcal{E}_{\beta}}\Big|_{\mathcal{U}} = [U^{(pp)}]^{-1}_{\nu\gamma,\nu'\gamma'} [Z_{\beta,\nu' \gamma'}]^* \approx \delta_{\beta \gamma}
 \frac{Z_\nu}{M(\w_\nu^2 - \w^2)},
 \end{equation} 
where $Z_{\nu}$ denotes the static Born effective charge of mode $\nu$, and $\w_\nu^2$ is the corresponding eigenvalue of the 
Born-Oppenheimer dynamical matrix.}

\blue{With the above results in hand, we are now able to seek an approximate form
for Eq.~\eqref{eq:me} by taking two further steps. First, we neglect the frozen-spin clamped-ion 
contribution, \blue{$\bar{\alpha} \simeq 0.01$ ps/m}, which is largely irrelevant on the scale of Figs.~\ref{fig:Cr2O3_sus}(c)-(d).
Second, we observe (as anticipated in Section~\ref{sec:Cr2O3_static_chi}) that the representation of $U_{H_\alpha a}$
on the basis of Eq.~\eqref{eq:cr2o3modes} has the following structure,
\begin{equation}
\begin{split}
U_{H_\alpha,u_{\nu \beta}} \approx 0, \qquad U_{H_\alpha,{n \beta}} \approx -2g_1 \delta_{n 3} \delta_{\alpha\beta}.
\end{split}
\end{equation}
where the prefactor accounts for the constant of proportionality that
relates ${\bf e}_3$ to the physical macroscopic magnetization.
(The spin-spin Berry curvature provides an accurate measure of
the angular momentum carried by a given local mode; the factor of $2$
has to do with the normalization of the ${\bf e}_j$ eigenvectors by $1/2$.)
Then the dynamical magnetoelectric susceptibility can be written as 
\begin{equation}
\alpha_{\alpha \beta}(\w) \approx -\frac{2g_1}{\Omega} \sum_{n\beta} \chi^{\rm RI}_{3\alpha,n\beta}(\w) \frac{\partial H_{n\alpha}(\w) }{\partial \mathcal{E}_\beta} \Big|_{\mathcal{U}}
\label{eq:me_model}
\end{equation}
where the relaxed-ion spin susceptibility matrix $\boldsymbol{\chi}^{\rm RI}$ is also expressed in the basis of Eq.~\eqref{eq:cr2o3modes}.} %

\blue{Based on the above analysis, the key physical quantities that govern the magnetoelectric
coupling in Cr$_2$O$_3$ are the clamped-ion Zeeman torques induced by either an electric field $\mathcal{E}_\beta$,
$H^{\mathcal{E}_{\beta}}_{n \alpha} (\w)$, or by the $\beta$ component of a given $E$ phonon irrep, $H^{u_{\nu\beta}}_{n\alpha}(\w)$.
Their frequency dependence can be accurately represented in terms of a static coupling plus a 
Berry curvature contribution,
\begin{equation}
H^{\lambda_\beta}_{n \alpha} (\w) = \frac{\partial H_{n\alpha}(\w)}{\partial \lambda_\beta} \Big|_{U} \simeq K_{{n \alpha },{\lambda_\beta}} + i \w G_{{n \alpha },{\lambda_\beta}},
\end{equation}
where $\lambda=\mathcal{E},u_\nu$.
In particular, $K$ and $G$ take the following form,
\begin{equation}
{\bf K}^{n \lambda} = \begin{pmatrix} 
0 \\ k^\lambda_2 \mathcal{A}  \\ k^\lambda_3 \mathcal{S} \\ 0 
\end{pmatrix}, \qquad {\bf G}^{n \lambda} = \begin{pmatrix} 
g^\lambda_1 \mathcal{A}  \\ 0 \\ 0 \\ g^\lambda_4 \mathcal{S}  
\end{pmatrix}.
\end{equation}
From the point of view of symmetry, these tensors have the same structure as ${\bf K}^{n3}$
and ${\bf G}^{n3}$ (the third columns of Eqs.~\eqref{eq:K_modes} and~\eqref{eq:G_modes} , respectively), which couple ${\bf e}_n$ with the ferromagnetic order parameter ${\bf e}_3$. 
In other words, the electric field and $E_u$ phonon perturbations
both transform like a macroscopic Zeeman field, a signature of the magnetoelectric effect.}

\blue{In the static regime, $k_3^{\lambda}$ describe a torque on ${\bf e}_3$ produced by an in-plane electric field or $E_u$
phonon. 
As such, they are the main physical parameters entering the 
static magnetoelectric coefficient of Cr$_2$O$_3$, which can be written (within the approximations
of this Section) as
\begin{equation}
\alpha_\perp^{\rm RI} \simeq -\frac{2g_1}{\Omega}  \chi_{33} \left( k^{\mathcal{E}}_3 +
\sum_\nu \frac{k^{\nu}_3 Z_\nu }{M_0 \w_\nu^2} \right);
\end{equation}
here $\chi_{33}=\chi^{\rm RI}_{3\alpha,3\alpha}(0) \simeq \chi^{\rm CI}_{3\alpha,3\alpha}(0)$ is the relevant component of the static spin susceptibility matrix.
The first term in the round brackets governs the clamped-ion response: we find $k^{\mathcal{E}}_3=-0.383$ meV/(V\AA$^{-1}$) in our strained geometry,
to be compared with $k^{\mathcal{E}}_3=-1.437$ meV/(V\AA$^{-1}$) in the experimental structure. (This large difference explains the underestimation of
$\alpha^{\rm CI}$ that we have pointed out earlier.)
The second term in the round brackets, accounting for the lattice-mediated contribution,
bears strong connections to the estaablished linear-response theory~\cite{IniguezPRL8}, 
with the main (cosmetic) difference that the magnetic charges of Ref.~\cite{IniguezPRL8} are expressed 
here in terms of the Zeeman torques $k^\nu_3$. (Our calculated values are reported in Table~\ref{tab:me_couplings}.)}

\begin{table}
\setlength{\tabcolsep}{7pt}
\begin{center}
\caption{Magnetoelectric coupling parameters for the E$_u$ phonon irreps of Cr$_2$O$_3$. From left to right: Born-Oppenheimer eigenfrequencies ($\w_\nu$) and effective charges ($Z_\nu$), and static ($k_3^\nu$) and dynamic ($g_1^\nu$) part of the torques induced on magnetic modes ${\bf e}_3$ and ${\bf e}_1$, respectively. }
\begin{tabular}{crrr}\hline\hline
   \T\B $\w_\nu$ (meV) & \multicolumn{1}{c}{$Z_\nu$ (e)} & \multicolumn{1}{c}{$k_3^\nu$} (meV/\AA) & \multicolumn{1}{c}{$g_1^\nu$ (10$^{-3}$/\AA)}  \\ \hline
   \T\B 36.2 & 1.55   &  $-$13.43  &  $-$5.58 \\
   \T\B 55.2 & 2.17   &  $-$0.23   &  $-$5.74 \\
   \T\B 62.9 & 22.16  &  $-$40.89  &  $-$22.47 \\
   \T\B 69.5 & 14.94  &      6.34  &  $-$13.28 \\
     \hline \hline
\end{tabular}
\label{tab:me_couplings}
\end{center}
\end{table}

\blue{In the dynamical regime, the rate of change of $\lambda$ produces, via the Berry curvature, additional
torques on ${\bf e}_1$ ($g_1^{\mathcal{E}}=-2.269\times10^{-3}$ \AA/V;
the values of $g^\nu_1$ corresponding to the
four $E_u$ irreps are reported in Table~\ref{tab:me_couplings})
and ${\bf e}_4$; to the best of our knowledge, these couplings have not been discussed before.
To assess their importance in the dynamical ME response, we  
perform a computational experiment where we set the Berry curvatures to zero
(i.e., we approximate $H^{\lambda_\beta}_{n\alpha } (\w) \rightarrow K_{{n\alpha},{\lambda_\beta}}$),
and recalculate the frequency-dependent ME response via Eq.~\eqref{eq:me_model} while keeping all the other
parameters unchanged.}

\begin{figure}
 \centering
 \includegraphics[width=1.0\columnwidth]{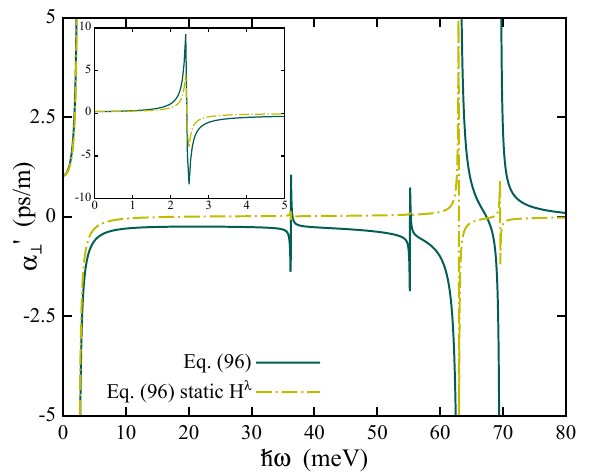}
 \caption{Real-part of the in-plane RI magnetoelectric coefficient calculated with Eq.~\eqref{eq:me_model} including (blue, solid) and neglecting (yellow, dash dot) dynamical contributions to the local Zeeman torques induced by electric fields and phonons. The inset shows a blow up of the CI acoustic magnon resonance.}
 \label{fig:Cr2O3_me_model}
\end{figure}

\blue{The results, shown in Fig.~\ref{fig:Cr2O3_me_model}, point to a remarkable impact of the Berry-curvature-related
couplings on all spectral features. 
On one hand, the characteristic structures that are associated with the phonon resonances  
appear strongly suppressed in amplitude and opposite in sign compared to the fully frequency-dependent 
solution; this indicates that the dynamical torque on ${\bf e}_1$, mediated by $g^\nu_1$, largely
dominates over the static torques due to $k^\nu_3$ in the phonon frequency range.
Such an outcome is especially remarkable when compared to the CrI$_3$ case discussed earlier, 
where the spin-phonon Berry curvature has a negligible impact on the calculated electromagnetic 
spectra.
Instead, in Cr$_2$O$_3$ $g^\nu_1$ is entirely responsible for the phase reversal
of $\alpha(\w)$ in a vicinity of the phonon resonances, pointing to a qualitative difference in  
how electromagnons behave in the two materials.
On the other hand, the amplitude of the acoustic magnon resonance appears 
significantly affected as well.
The most dramatic effect occurs at the clamped-ion level, where the neglect of ${\bf G}^{1\mathcal{E}}$
leads to a suppression of the corresponding spectral feature by a factor of 2. (At the RI level the
neglect of both ${\bf G}^{1\mathcal{E}}$ and ${\bf G}^{1\nu}$ entails a 20\% suppression.) 
The same Berry curvature is also responsible for the persistence of a nonvanishing magnetoelectric response
far away from the resonance, where 
{$\alpha^{\rm CI}(\w \gg \w_{\rm AM}) \approx g_1^\mathcal{E}/g_{1}$}.}

\begin{figure}
 \centering
 \includegraphics[width=1.0\columnwidth]{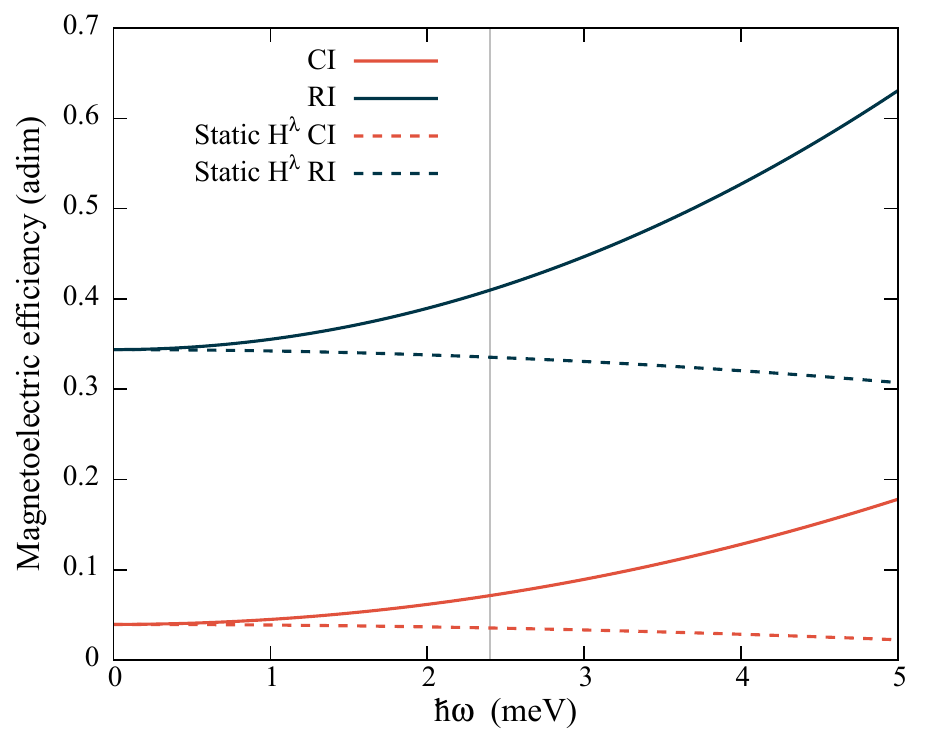}
 \caption{Magnetoelectric efficiency, as defined in Eq.~\eqref{eq:me_eff}. Results are reported at the CI (red) and RI (blue) levels, with (solid lines) and without (dashed lines) inclusion of the dynamical contributions to the local Zeeman torques induced by electric fields and phonons. The vertical grey line indicates the position of the AM resonance.}
 \label{fig:me_eff}
\end{figure}

\blue{With the above results in hand, we can directly compare the coupling strength between the acoustic magnon
and a macroscopic Zeeman or electric field; this can be regarded as a
 first-principles realization of the experiment of Bilyk et al.~\cite{bilyk2025l}.
To that end, in Fig.~\ref{fig:me_eff} we plot the \emph{magnetoelectric efficiency}, which we define as
the ratio of the dynamical magnetoelectric response (in dimensionless units of $1/c$)
and magnetic susceptibility,
\begin{equation}
\gamma^{\rm ME}(\w) = \frac{c \alpha(\w)}{\chi(\w)}.
\label{eq:me_eff}
\end{equation}
At the acoustic magnon resonance, our full DFPT description yields $\gamma^{\rm ME}\simeq 0.4$, in fair agreement with
the experimental findings of Ref.~\cite{bilyk2025l}, which indicate a ratio of the order of unity. 
Note the drastic enhancement (roughly a factor of 20) of $\gamma^{\rm ME}$ that is associated with the inclusion of lattice-mediated contributions; a neglect of the latter would lead to a marked discrepancy with the results of Ref.~\cite{bilyk2025l}.
Also the Berry curvatures yield a significant enhancement of the dynamical magnetoelectric coupling coefficient compared to 
the static one, consistent with the discussion of the previous Section.}

\section{Conclusions}

In summary, we have presented a general methodological framework to calculate the
dynamical response of noncollinear magnets to arbitrary external perturbations.
The foundations of our theory rest on a general result: the equivalence of the
penalty-function and Lagrange-multiplier approaches to constrained DFT.
Establishing their mutual relation via an exact mapping in parameter space,
which we express in the language of Legendre transforms, can be regarded as
one of the main formal achievements of this work.
Our strategy overcomes several long-standing difficulties of existing first-principles approaches,
related to the poor conditioning of the magnetic degrees of freedom in the calculations, and to the
lack of an exact linear-response framework to treat the coupled \emph{nonadiabatic} dynamics of
phonons and spins.
Thanks to the methods developed here, one can now perform DFPT calculations in magnets with
comparable efficiency, accuracy and ease as in standard nonmagnetic insulators.

As a first application, we have tested the adiabatic approach to the spin~\cite{NiuPRL98,lin2025} and coupled
spin-phonon~\cite{RenPRX24} problem that were proposed in earlier works. While we find good numerical
agreement with our full frequency-dependent treatment in \blue{both CrI$_3$ and Cr$_2$O$_3$}, we propose
a procedure to systematically improve its accuracy, which we refer to as ``second-order
adiabatic approximation''.
Our main formal result here consists in the demonstration that magnons are endowed with a
nonzero mass, originating from the inertia of the electrons that are dragged along during spin
precession.
This result calls for a revision of the Landau-Lifshitz equations, via the inclusion of an
effective mass term at second order in the frequency.

As a second application, we have demonstrated the calculation of the
\blue{frequency-dependent dielectric, magnetic and magnetoelectric susceptibilities} of the system,
including the contributions from the
coupled spin-lattice dynamics in the external electric field.
In addition to recovering several known results (e.g., the partial shift of
spectral weight from optical phonon to magnon resonances),
we have identified some interesting new features; for example,
some general properties of hybrid phonon-electromagnon
modes that were formerly unknown or overlooked.
Of particular note, we generalize the spin-phonon model of Ref.~\onlinecite{RenPRX24}
by introducing the coupling to the external electric field. This allows us to
understand, to a high degree of accuracy, our full first-principles results
in terms of simple analytic formulas that depend on a minimal set of parameters.

As an outlook, our work opens countless opportunities for future research.
Our approach nicely complements the
adiabatic theory of Ref.~\cite{RenPRX24}, by providing an efficient DFPT route to calculating
the required Hessians and Berry curvatures.
A natural next step will consist in combining the two approaches to enable
Fourier interpolation of the coupled magnon and phonon bands over the
whole Brillouin zone; work towards this goal is under way.
Another obvious future goal consists in achieving a better integration of
our linear-response method with the preexisting algorithms to calculate
the electronic ground state at constrained magnetic moments.
This task will enable calculating second
derivatives in an arbitrary spin configuration, which
is especially interesting in the perspective of constructing
``second-principles'' spin models.
Also, one could incorporate the ideas of Ref.~\onlinecite{lin2025} to go beyond the limitations
of the atomic-sphere approximation. 
Last but not least, it will be interesting to combine the formalism presented here
with the recently developed first-principles approach to spatial dispersion~\cite{Royo2019}.
This appears as a very timely addition to address the pressing experimental
questions in the context of optical activity~\cite{ZabaloPRL23}, where magnons can
play a key role~\cite{gao2024}.
Also flexomagnetism~\cite{EdstromPRL22} (describing the modification of the spin
couplings due to strain gradients) can be regarded as a spatial dispersion effect
that has received increasing attention as of late, but has been largely out
of reach in the context of first-principles methods.
The possibility of systematically calculating flexomagnetic coefficients via
DFPT calculations on the primitive crystal cell would open many exciting
research avenues in this area.

\begin{acknowledgments}
We thank Shang Ren and David Vanderbilt for many illuminating discussions and for carefully testing our code.
MR thanks the Center for Materials Theory at Rutgers University for its kind hospitality and stimulating research environment.
This work was supported by the State Investigation Agency through the Severo Ochoa Programme for Centres of Excellence in R\&D (CEX2023-001263-S), by the Ministry of Science, Innovation and Universities (Grants No. PID2023-152710NB-I00 and No. PRX22/00390) and from Generalitat de Catalunya (Grant No. 2021 SGR 01519).
\end{acknowledgments}

\appendix

\section{Minimal model of CrI$_3$ \label{sec:model}}

To understand the results presented so far, we can reason in terms of a minimal model of
the coupling between the $E_u^+$ phonon, the optical magnon and the (circularly polarized)
electric field $\mathcal{E}^+$.
All the relevant coupling coefficients are then contained in a $3\times 3$ Hermitian matrix,
consisting in the second derivatives of $U(\tau,M,\mathcal{E})$ within the aforementioned
subspace of parameters.
Their values are readily obtained via projection of the full ${\bf U}$ matrix on the relevant mode vectors.
For the optical magnon and electric field we use the ${\bf e}^{\rm OM}_+$ and ${\bf e}^\mathcal{E}_+$
components introduced in Sections~\ref{sec:locmagsus} and~\ref{sec:dieltens}.
Regarding the phonon, we define the \emph{eigendisplacement} vectors associated with the $E_u$ 
doublet as
\begin{equation}
e^{\rm ph}_{\pm, \kappa \alpha} = v^{\pm}_{\kappa\alpha} \sqrt{\frac{M_0}{M_\kappa}},
\end{equation}
where ${\bf  v}^{\pm}$ are circularly polarized eigenvectors of the static dynamical matrix,
and $M_0$ is an arbitrary mass factor, whose purpose is to guarantee that the phonon amplitude 
$\tau$ has a physical dimension of length.

Based on our results of Section~\ref{sec:lattdyn}, we can write the (inverse of the) bare phonon propagator, to a very good accuracy,
in terms of the Born-Oppenheimer frequency,
\begin{equation}
U^{(pp)}_{++} =  ({\bf e}^{\rm ph}_+)^* \cdot {\bf U}^{(pp)} \cdot {\bf e}^{\rm ph}_+ \approx M_0 (\w_0^2 - \w^2).
\end{equation}
Next, we calculate the spin-phonon coupling coefficient as
\begin{equation}
H^\tau_{\pm} = ({\bf e}^{\rm OM}_\pm)^* \cdot {\bf U}^{(sp)} \cdot {\bf e}^{\rm ph}_\pm,
\end{equation}
with 
the physical meaning of the torque on the OM induced by the lattice distortion.
We find that $H^\tau_{\pm}(\w)$ has a weak frequency dependence and is well approximated by its static value, which is $H^\tau_{\pm}(0)=(-8.347 \mp i70.163)$ meV/\AA~  according to our calculations.
Finally, we define the Born dynamical charge of the $E_u$ mode as
\begin{equation}
Z_+ =  -({\bf e}^\mathcal{E}_+)^* \cdot {\bf U}^{(\mathcal{E} p)} \cdot {\bf e}^{\rm ph}_+,
\end{equation}
which we approximate with its static limit where it becomes a real constant $Z=15.816$ $|\rm e|$ .

By combining the above definitions with the static limit of the parameter $H^\mathcal{E}_\pm$ introduced in Section~\ref{sec:dieltens}, $H^\mathcal{E}_\pm(0)=(-0.810 \mp i1.097)$ meV/(V\AA$^{-1}$), we can provide a simple analytical representation that reproduces faithfully most of our first-principles results.
For example, we write the nonadiabatic (relaxed-spin) phonon propagator of Eq.~\eqref{fcal_pp} as the inverse of the following scalar function, 
\begin{equation}
F^{(pp)}_{\pm \pm} \approx M_0 (\w_0^2 - \w^2) - \frac{|H^\tau|^2}{K \mp \w G - \w^2 M}.
\end{equation}
Then, the poles of the propagator are given by the zeros of its denominator,
\begin{equation}
\label{zeros}
(\w_0^2 - \w^2) (K \mp \w G - \w^2 M) - \frac{|H^\tau|^2}{M_0} =0.
\end{equation}
This equation has three real roots, yielding the frequencies of the hybrid phonon-magnon modes derived from $E_u^+$, $E_u^-$ and OM.
If we neglect the effective magnon mass parameter, $M$, Eq.~\eqref{zeros} recovers the minimal model of Ref.~\onlinecite{BoniniPRL23}.

Also, starting from Eq.~\eqref{fcal_ss} we can obtain the following approximate formula for the optical magnon channel of $\chi^{\rm RI}$,
\begin{equation}
\label{chi_approx}
\chi^{\rm RI}_{\pm\pm} \approx \frac{(\w_0^2 - \w^2)}{(\w_0^2 - \w^2) (K \mp \w G - \w^2 M) - \frac{|H^\tau|^2}{M_0}}.
\end{equation}
This expression shares the exact same denominator as the nonadiabatic phonon propagator, and 
yields a result for $\chi^{\rm RI}(\w)$ that is indistinguishable from the fully first-principles result on the scale of Fig.~\ref{fig:magsus}.

Next, our results for the full frequency-dependent dielectric function 
can be understood by approximating the contribution of the coupled spin-lattice system as
\begin{equation}
\label{eps_approx}
\begin{split}
\Delta \epsilon^{\rm RI} \approx& \frac{4\pi}{\Omega} \frac{|Z|^2}{F^{(pp)}(\w)},
\end{split}
\end{equation}
where we have neglected the (small) direct coupling between the field and the spins.
Eq.~\eqref{eps_approx} yields a description of the RI/RS dielectric tensor that matches the full first-principles curves of Fig.~\ref{fig:dielectric} to a very good accuracy, confirming that the IR intensity of the OM-like hybrid indeed originates from its phonon fraction.

Finally, in a vicinity of the coupled $E_u$--OM resonances, we can approximate Eq.~\eqref{eq:emag_ri} as
\begin{equation}
\frac{\partial m_+}{\partial \mathcal{E}_+} \simeq -\chi_{++}^{\rm RI} \frac{H^\tau_+ Z}{M_0 (\omega_0^2 - \omega^2)}.
\end{equation}
By using Eq.~\eqref{chi_approx} for the RI susceptibility,
we obtain 
\begin{equation}
\frac{\partial m_+}{\partial \mathcal{E}_+} \approx -\frac{C}{(K-\w G -\w^2 M)(\w_u^2 - \w^2) - \frac{|H^\tau|^2}{M_0}},
\label{eq:ame_model}
\end{equation}
where $C=H^\tau_+ Z/M_0$ is a frequency-independent complex number that is responsible for the phase delay via $H^\tau_+$, 
and the denominator explains the equally large (but with opposite sign) spin response at the phonon- and magnon-like resonance frequencies.

\bibliography{merged}

\end{document}